\documentclass[fleqn,usenatbib]{mnras}

\usepackage{newtxtext,newtxmath}

\usepackage[T1]{fontenc}

\DeclareRobustCommand{\VAN}[3]{#2}
\let\VANthebibliography\thebibliography
\def\thebibliography{\DeclareRobustCommand{\VAN}[3]{##3}\VANthebibliography}

\usepackage{graphicx}	
\usepackage{amsmath}	

\usepackage{graphicx}	
\usepackage{amsmath}	
\usepackage{xspace}
\usepackage{longtable}
\usepackage{subfig}
\usepackage{booktabs}
\usepackage{lscape}
\usepackage{siunitx}
\usepackage[symbol]{footmisc}
\usepackage[dvipsnames]{xcolor}
\usepackage{hyperref}

\usepackage{tabularx}

\newcommand{\radiusA}{\,$R_\mathrm{b}=0.98\pm0.07\,\mathrm{R}_\oplus$ } 
\newcommand{\periodA}{\,$P=4.7185898^{+0.0000054}_{-0.0000041}\,\mathrm{d}\xspace$ } 
\newcommand{\instellationA}{\,$S_\mathrm{b}=4.4\pm1.1\,\mathrm{S}_\oplus$\xspace} 

\newcommand{\radiusB}{\,$R_\mathrm{b}=3.56\pm0.21\,\mathrm{R}_\oplus$\xspace } 
\newcommand{\periodB}{\,$P=6.2340258^{+0.0000034}_{-0.0000036}\,\mathrm{d}$ } 
\newcommand{\instellationB}{\,$S_\mathrm{b}=4.9\pm1.1\,\mathrm{S}_\oplus$\xspace } 

\title[TOI-6716\,b \& TOI-7384\,b; temperate planets]{\textcolor{black}{Two temperate Earth- and Neptune-sized planets orbiting fully convective M dwarfs}}

\author[M. G. Scott \& G. Dransfield et al.]{
Madison G. Scott$^{\href{https://orcid.org/0009-0006-3846-4558}{\includegraphics[scale=0.5]{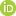}}}$,$^{1}$\thanks{E-mail: mgs947@student.bham.ac.uk (MGS); george.dransfield@magd.ox.ac.uk (GD)}\thanks{ These two authors contributed equally to this work and should be considered joint first authors.}
Georgina Dransfield$^{\href{https://orcid.org/0000-0002-3937-630X}{\includegraphics[scale=0.5]{figures/orcid.jpg}}}$,$^{2,3,1}$\color{blue}$^{\star}{\dagger}$
\color{black}
Mathilde Timmermans$^{\href{https://orcid.org/0009-0008-2214-5039}{\includegraphics[scale=0.5]{figures/orcid.jpg}}}$,$^{1, \text{4}}$, 
Amaury H.M.J. Triaud$^{\href{https://orcid.org/0000-0002-5510-8751}{\includegraphics[scale=0.5]{figures/orcid.jpg}}}$,$^{1}$ 
\newauthor
Benjamin V. Rackham$^{\href{https://orcid.org/0000-0002-3627-1676}{\includegraphics[scale=0.5]{figures/orcid.jpg}}}$,$^{\text{5}, \text{6}}$ 
Khalid Barkaoui$^{\href{https://orcid.org/0000-0003-1464-9276}{\includegraphics[scale=0.5]{figures/orcid.jpg}}}$,$^{\text{7}, \text{4}, \text{5}}$ 
Adam J. Burgasser$^{\href{https://orcid.org/0000-0002-6523-9536}{\includegraphics[scale=0.5]{figures/orcid.jpg}}}$,$^{\text{8}}$ 
Karen A. Collins$^{\href{https://orcid.org/0000-0001-6588-9574}{\includegraphics[scale=0.5]{figures/orcid.jpg}}}$,$^{\text{9}}$ 
\newauthor
Michaël Gillon$^{\href{https://orcid.org/0000-0003-1462-7739}{\includegraphics[scale=0.5]{figures/orcid.jpg}}}$,$^{\text{4}}$ 
Steve B. Howell$^{\href{https://orcid.org/0000-0002-2532-2853}{\includegraphics[scale=0.5]{figures/orcid.jpg}}}$,$^{\text{10}}$ 
Alan M. Levine$^{\href{https://orcid.org/0001-8172-0453}{\includegraphics[scale=0.5]{figures/orcid.jpg}}}$, $^{\text{6}}$ 
Francisco J. Pozuelos$^{\href{https://orcid.org/0000-0003-1572-7707}{\includegraphics[scale=0.5]{figures/orcid.jpg}}}$,$^{\text{11}}$ 
\newauthor
Keivan G. Stassun$^{\href{https://orcid.org/0000-0002-3481-9052}{\includegraphics[scale=0.5]{figures/orcid.jpg}}}$,$^{\text{12}}$ 
Carl Ziegler$^{\href{https://orcid.org/0000-0002-0619-7639}{\includegraphics[scale=0.5]{figures/orcid.jpg}}}$,$^{\text{13}}$ 
Yilen Gomez Maqueo Chew$^{\href{https://orcid.org/0000-0002-7486-6726}{\includegraphics[scale=0.5]{figures/orcid.jpg}}}$,$^{\text{14}}$ 
Catherine A. Clark$^{\href{https://orcid.org/0000-0002-2361-5812}{\includegraphics[scale=0.5]{figures/orcid.jpg}}}$,$^{\text{15}}$ 
\newauthor
Yasmin Davis$^{\href{https://orcid.org/0009-0000-6625-137X}{\includegraphics[scale=0.5]{figures/orcid.jpg}}}$,$^{1}$ 
Fatemeh Davoudi$^{\href{https://orcid.org/0000-0002-1787-3444}{\includegraphics[scale=0.5]{figures/orcid.jpg}}}$,$^{\text{4}}$ 
Tansu Daylan$^{\href{https://orcid.org/0000-0002-6939-9211}{\includegraphics[scale=0.5]{figures/orcid.jpg}}}$,$^{\text{16}}$ 
Brice-Olivier Demory$^{\href{https://orcid.org/0000-0002-9355-5165}{\includegraphics[scale=0.5]{figures/orcid.jpg}}}$,$^{\text{17}}$
\newauthor
Dax Feliz$^{\href{https://orcid.org/0000-0002-2457-7889}{\includegraphics[scale=0.5]{figures/orcid.jpg}}}$,$^{\text{18}}$ 
Akihiko Fukui$^{\href{https://orcid.org/0000-0002-4909-5763}{\includegraphics[scale=0.5]{figures/orcid.jpg}}}$,$^{\text{19}, \text{7}}$ 
Maximilian~N.~Günther$^{\href{https://orcid.org/0000-0002-3164-9086}{\includegraphics[scale=0.5]{figures/orcid.jpg}}}$,$^{\text{20}}$
Emmanuël Jehin,$^{\text{21}}$
\newauthor
Florian Lienhard$^{\href{https://orcid.org/0000-0003-4047-0771}{\includegraphics[scale=0.5]{figures/orcid.jpg}}}$,$^{\text{22}}$ 
Andrew W. Mann$^{\href{https://orcid.org/0000-0003-3654-1602}{\includegraphics[scale=0.5]{figures/orcid.jpg}}}$,$^{\text{23}}$ 
Clàudia Janó Muñoz$^{\href{https://orcid.org/0009-0008-5713-0750}{\includegraphics[scale=0.5]{figures/orcid.jpg}}}$,$^{\text{24}}$ 
Norio Narita$^{\href{https://orcid.org/0000-0001-8511-2981}{\includegraphics[scale=0.5]{figures/orcid.jpg}}}$,$^{\text{19}, \text{25}, \text{7}}$ 
\newauthor
Peter P. Pedersen,$^{\text{24}, \text{22}}$
Richard P. Schwarz$^{\href{https://orcid.org/0000-0001-8227-1020}{\includegraphics[scale=0.5]{figures/orcid.jpg}}}$,$^{\text{9}}$ 
Avi Shporer$^{\href{https://orcid.org/0000-0002-1836-3120}{\includegraphics[scale=0.5]{figures/orcid.jpg}}}$,$^{\text{6}}$ 
Abderahmane Soubkiou$^{\href{https://orcid.org/0000-0002-0345-2147}{\includegraphics[scale=0.5]{figures/orcid.jpg}}}$,$^{\text{4}}$ 
\newauthor
Sebastián Zúñiga-Fernández$^{\href{https://orcid.org/0000-0002-9350-830X}{\includegraphics[scale=0.5]{figures/orcid.jpg}}}$,$^{\text{4}}$
\\
Affiliations are listed at the end of the paper.
}

\date{Accepted XXX. Received YYY; in original form ZZZ}

\pubyear{\the\year{}}

\begin{document}
\label{firstpage}
\pagerange{\pageref{firstpage}--\pageref{lastpage}}
\maketitle

\begin{abstract}
As the diversity of exoplanets continues to grow, it is important to revisit assumptions about habitability and classical HZ definitions. In this work, we introduce an expanded 'temperate' zone, defined by instellation fluxes between $0.1<S/\mathrm{S}_\oplus<5$, thus encompassing a broader range of potentially habitable worlds. We also introduce the TEMPOS survey, which aims to produce a catalogue of precise radii for temperate planets orbiting M dwarfs with $T_\mathrm{eff}\leq3400\,$K. This work reports the discovery and characterisation of two planets in this temperate regime \textcolor{black}{orbiting mid-type M dwarfs}: TOI-6716\,b, a \radiusA planet orbiting its M4 host star ($R_\star=0.231\,\pm0.015\mathrm{R}\,_\odot$, $M_\star=0.223\pm0.011\,\mathrm{M}\,_\odot$, $T_\mathrm{eff}=3110\pm80\,\mathrm{K}$) with a period \periodA, and TOI-7384\,b, a \radiusB planet orbiting an M4 ($R_\star=0.319\,\pm0.018\mathrm{R}\,_\odot$, $M_\star=0.318\pm0.016\,\mathrm{M}\,_\odot$, $T_\mathrm{eff}=3185\pm75\,\mathrm{K}$) star every \periodB. The radii of TOI-6716\,b and TOI-7384\,b have precisions of $6.8\%$ and $5.9\%$ respectively. We validate these planets with multi-band ground-based photometric observations, high-resolution imaging and statistical analyses. We find these planets to have instellation fluxes close to the inner (hotter) edge of the temperate zone, with \instellationA and \instellationB for TOI-6716\,b and TOI-7384\,b respectively. \textcolor{black}{Also, with a predicted TSM similar to the TRAPPIST-1 planets, TOI-6716\,b is likely to be a good rocky-world JWST target, should it have retained its atmosphere.}
\end{abstract}




\section{Introduction} \label{sec:intro}
\color{black}
The search for transiting temperate planets, where for the purpose of this paper we define `temperate' as a planet with instellation flux $0.1\leq S/\mathrm{S}_\oplus \leq 5$, around solar-like stars is notoriously challenging. This is because planet equilibrium temperature scales with both semi-major axis and stellar effective temperature, leading to low transit probabilities. However, thanks to the lower effective temperatures of M-type stars, and in particular mid- to late-type M dwarfs ($\leq3400\,$K), temperate planets orbiting such stars are far more likely to transit, and therefore become more accessible in transit surveys \citep[e.g.,][]{Delrez2022, Dransfield2023, dholakia_palethorpe2024}. Additionally, due to the relatively small sizes of the M dwarf host stars, Earth-sized temperate planets produce deeper and more detectable transit signals than a similar planet orbiting a solar-type star (e.g., depths $\sim 0.3-13\,\rm ppt$ for the M dwarf radius range $0.08\leq R_\star / \mathrm{R}_\odot \leq 0.5$, compared with $\sim 0.1\,\rm ppt$ for a $R_\star=1\mathrm{R}_\odot$ star). This allows for more precise constraints on parameters such as planetary radius, inclination and orbital period from photometric data using both space-based \citep[e.g., \textit{TESS}; ][]{ricker2015} and ground-based telescopes \citep[e.g., SPECULOOS; ][]{speculoos}. A further benefit is that the measurement of precise planetary masses of smaller planets via radial velocities may be facilitated by the higher planet-to-star mass ratios \citep[see e.g., ][]{lhs1140b, turner2025}.

To date, 281 temperate planets have been confirmed (90 transiting\footnote{NASA exoplanet archive, June 2025; \url{https://exoplanetarchive.ipac.caltech.edu}. Recalculated from quoted stellar luminosities and semi-major axes.}), with just less than half (118) orbiting M dwarfs ($\leq4000\,$K).
Of these, 39 have been identified by \textit{TESS} \citep[see, e.g.,][]{cadieux2022, murgas2023, timmermans2024}. Temperate planets account for only 2.6\% of all confirmed planets to date; this is unsurprising given most known systems are hosted by FGK stars and transit survey missions to date have had limited sensitivity to long orbital periods. 
For example, typical \textit{TESS} observations of targets outside the continuous viewing zone span only $\sim$27 days every 2 years; we calculate that for a temperate planet around an FGK-type star, the expected orbital periods would be in the range of $\sim$21\,d ($\sim$K9V, $T_\mathrm{eff}\simeq3930\,$K) to $\sim$7200\,d ($\sim$F0V, $T_\mathrm{eff}\simeq7220\,$K ). Even the shortest of these periods often result in, at most, a single observed transit per sector, making them difficult to identify as a planet candidate. In contrast, for mid- to late-type M dwarfs, the corresponding period range shifts from 0.9\,d ($\sim$M9V, $T_\mathrm{eff}\simeq2380\,$K) to 147\,d ($\sim$M3V, $T_\mathrm{eff}=3400\,$K), bringing a much larger fraction of the temperate planet population within the detectability window of \textit{TESS} and other transit surveys. Consequently, late-type M dwarfs represent a favourable stellar population for identifying temperate-zone planets. A key example of this is the temperate TRAPPIST-1 planets, whose orbital periods lie in the range $1.5-18.8\,$d with instellation fluxes in the range $0.13\lesssim S/\mathrm{S}_\oplus \lesssim4.3$ \citep{trappist1}.

\color{black}

A `temperate zone' has not formally been defined in the literature, but its value lies in broadening current ideas of Earth-centric habitability. This approach expands the classical HZ described by \citet{kasting1993} and \citet{kopparapu2014}. Planets that may fall outside this traditional metric, such as Hycean or water worlds \citep{Madhu2021} and temperate sub-Neptunes \citep{seager2021}, could contain environments that satisfy a range of habitability requirements which at present remain poorly understood. By broadening parameters associated with habitability, a wider range of planetary conditions can be encompassed, thus enabling more diverse interpretations of where life might exist beyond Earth.

\color{black}
In the era of \textit{JWST} \citep{jwst2006, gardner2023, jwstERS}, feasible exoplanet atmospheric characterisation observations are limited not only by the amplitude of atmospheric features, by also by the frequency with which repeated observations can be obtained. The low stellar luminosities of mid-to-late-type M dwarfs place their temperate zones at short orbtial periods, and their small radii allow for relatively deep transits even from small planets \citep[see, e.g.,][]{triaud2013_bds}. These factors make temperate planets around nearby, bright, late-type M dwarfs among the most favourable \textit{JWST} targets, allowing for high signal-to-noise ratio (SNR) detections of key atmospheric features with only a few hours of telescope time \citep{CharbonneauDeming2007, Morley2017}.

\color{black}
This paper increases this small, but growing sample of planets with both Earth and Neptune-sized exoplanets orbiting mid-to-late-type M dwarfs. While these planets are not in the HZ, they sit at the inner (hotter) edge of the temperate zone, facilitating studies of how different types of warm planets form and evolve around fully-convective M dwarfs, i.e, in systems with environments much different from the environment in our own solar system. In this work we also introduce a survey designed to produce a catalogue of precise radii for temperate planets orbiting late-type M ($\leq3400\,$K) dwarfs using the SPECULOOS \citep[Search for habitable Planets EClipsing ULtra-cOOl Stars;][]{speculoos,ZunigaFernandez2025}, described in Section~\ref{sec:tempos}.

\textcolor{black}{Our paper is organised as follows: firstly, we start with defining the temperate zone in Section~\ref{sec:tempzone}, followed by an introduction to the TEMPOS survey in Section~\ref{sec:tempos}. In Section~\ref{sec:stellar_char}, we characterise the host stars using reconnaissance spectroscopy and their spectral energy distributions. Next, we describe the identification of the planet candidates from \textit{TESS} data in Section~\ref{sec:planet_identification}. Section~\ref{sec:vettingvalidation} outlines the methods to validate the planet candidates, and describes our ground-based follow-up efforts that contribute to this. We then describe the global analysis of all photometric data for each planet in Section~\ref{sec:analysis}, followed by a discussion and conclusion of our results in Section~\ref{sec:discussion}.}

\color{black}

\section{Definition of the `temperate zone'} \label{sec:tempzone}

Throughout the exoplanet literature, especially in the context of small and potentially habitable planets, we encounter the word `temperate'. In general, each work makes it clear how they define this term. For instance, \cite{Greklek2025} presenting the TOI-1266 system, define `temperate' to mean having an equilibrium below 450~K. This definition includes both transiting planets in this system, the inner of which has T$_{\rm eq}=415$~K and instellation flux of $4.72~\rm S_{\oplus}$. However, \cite{Yang2024} define `temperate sub-Neptunes' as those having T$_{\rm eq}\lesssim 500$~K. `Temperate' is stated to mean an equilibrium temperature of 400~K in \cite{Encrenaz2022} and \cite{Piette2022}, while in \cite{Peterson2023} it is defined as $<400$~K. In this latter work, the `temperate' planet LP~791-18~d is presented, with T$_{\rm eq}=395.5$~K and instellation flux of $5.83~\rm S_{\oplus}$. In \cite{gunther2019}: TOI-270~d is presented as a `temperate' $2.13\rm ~R_{\oplus}$ planet with a T$_{\rm eq}=387$~K and an instellation flux of $3.92~\rm S_{\oplus}$. Here `temperate' is defined as the having equilibrium temperatures between the survival temperature for extremophiles ($395$~K) and the freezing point of water ($273.15$~K).

We also find that several works describe a planet as `temperate' without giving a specific definition. Examples include HD~35843~c, a `temperate' $2.54\rm ~R_{\oplus}$ planet with a T$_{\rm eq}=479$~K and an instellation flux of $11.38~\rm S_{\oplus}$ \citep{Hesse2025}, while TOI-2285~b is reported as a `venus-zone temperate' planet with $1.77\rm ~R_{\oplus}$ and T$_{\rm eq}=358$~K and instellation flux of $3.91~\rm S_{\oplus}$ \citep{Fukui2025,Miles2025}. Additionally, \cite{Mallorqu2023} present the young `temperate' mini-Neptune TOI-1801~b with a T$_{\rm eq}=493$~K and instellation flux of $10.73~\rm S_{\oplus}$. These planets have similar equilibrium temperatures, but instellation fluxes ranging from $3.91-11.4~\rm S_{\oplus}$, putting them beyond classical habitability arguments found in works like \cite{kopparapu2014}. 

Some works define a `temperate zone' in terms of stellar irradiation, as is done for the so-called `habitable zone'. For instance \cite{Cowan2015} define `temperate terrestrial planet' as one having $0.5\rm ~R_{\oplus} < R < 1.5\rm ~R_{\oplus}$ and $0.5\rm ~S_{\oplus} < S < 1.5\rm ~S_{\oplus}$. On the other hand, \cite{seager2021} define `temperate' simply to mean receiving `Earth-like' irradiation flux from a host star, as this could lead to liquid water in the atmosphere or on the surface. \cite{Triaud2024} again give a specific instellation range, defining `temperate' in their work to mean $0.25\rm ~S_{\oplus} < S < 4\rm ~S_{\oplus}$. We also find that in some works, like \cite{Reiners2018} and \cite{Sun2025}, `temperate zone' is used interchangeably with `habitable zone'. \cite{Wells2019} present K2-133~e as a `temperate zone' planet with a T$_{\rm eq}=296$~K and insolation flux of $1.8~\rm S_{\oplus}$. They note that the instellation puts it outside the HZ boundaries but they use the Earth Similarity Index \citep[ESI.;][]{Schulze2011} to assess `habitability' instead. The ESI is part of a two-tier system to quantify how Earth-like a planet is in terms of its size, density, escape velocity and mean surface temperature. 

And of course, all seven planets in the TRAPPIST-1 system are presented as `temperate', with equilibrium temperatures $172-398$\,K and instellation fluxes of $0.14-4.15~S_{\oplus}$ \citep{gillon2018}. For the inner planets, arguments about potential habitability are centred around the possibility of a localised habitable region on the planets, potentially at the terminator of these tidally locked worlds \citep{Menou2013}.

\begin{figure}
    \centering
    \includegraphics[width=1\columnwidth]{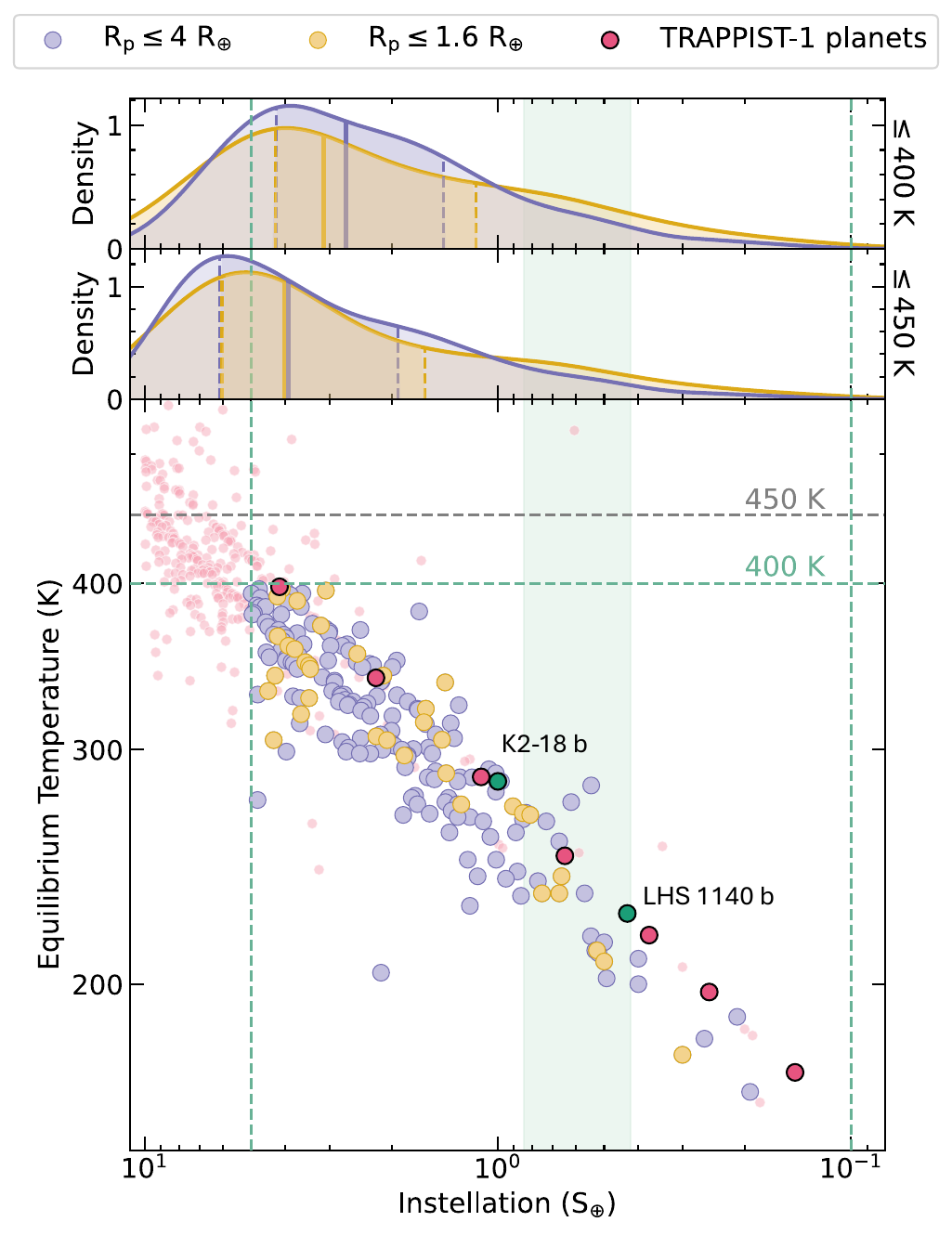}
    \caption{Our adopted definition of the temperate zone. In the lower panel we plot transiting planets retrieved from the NASA Exoplanet Archive 
    in terms of equilibrium temperature and instellation. Pink points in the foreground represent the full sample, purple points have $R_{\rm p}\leq 4~\rm R_\oplus$, and brown points have $R_{\rm p}\leq 1.5~\rm R_\oplus$. The seven temperate planets of the TRAPPIST-1 system are shown as red points, and in green we highlight K2-18~b and LHS 1140~b, two canonical temperate planets. The green shaded area indicates the conservative habitable zone as defined by \citealt{kopparapu2014}; the temperate zone is enclosed by the green dashed lines. The two upper panes show the KDE curves for the planet samples, with the colours matching the points in the lower panel.
    }
    \label{fig:temperate}
\end{figure}

For our work, we seek to formalise the definition of a `temperate zone' in terms of an instellation range, in a manner that incorporates as many as possible of the above definitions. We queried the NASA Exoplanet Archive for confirmed transiting exoplanets from the Composite Data table having irradiation fluxes $<10~S_{\oplus}$. These are plotted in Fig. \ref{fig:temperate}. As 400~K and 450~K are the two most common cut-offs we find in the literature for calling a planet `temperate', we perform a kernel density estimation (KDE) analysis of the instellation flux distribution, restricted to planets with $R_{\rm p}\leq 4 \rm R_{\oplus}$ and equilibrium temperatures either $\leq 400$~K or $\leq450$~K. In each case, we assess the distribution of all planets in the sample, as well as a sample that is further restricted to planets smaller than $1.5 \rm R_{\oplus}$ to see any differences for planets most likely to be rocky. The KDE curves are plotted in the upper panels of Fig. \ref{fig:temperate}.

For the $\leq 400$~K, we find median and 75$^{\rm th}$ percentile values of $2.687~\rm S_{\oplus}$ and $4.242~\rm S_{\oplus}$ respectively for the $R_{\rm p}\leq 4 \rm~ R_{\oplus}$ distribution, and $2.214~\rm S_{\oplus}$ and $4.205~\rm S_{\oplus}$ respectively for the $R_{\rm p}\leq 1.5 \rm ~R_{\oplus}$ distribution. These values are highlighted on the upper panel of Fig. \ref{fig:temperate}. The corresponding values for median and 75$^{\rm th}$ percentile for the $\leq 450$~K sample are $3.914S_{\oplus}$ and $6.1425~\rm S_{\oplus}$ respectively for the $R_{\rm p}\leq 4 \rm ~R_{\oplus}$ distribution, and $4.09~\rm S_{\oplus}$ and $6.1~\rm S_{\oplus}$ respectively for the $R_{\rm p}\leq 1.5~ \rm R_{\oplus}$ distribution. We highlight these values in the middle panel of Fig. \ref{fig:temperate}. 

We therefore place our upper boundary of the temperate zone at $S_{\rm p}=5 ~\rm S_{\oplus}$, just below the median of the 75$^{\rm th}$ percentiles ($5.17 ~\rm S_{\oplus}$). \textcolor{black}{By using the 75$^{\rm th}$ percentiles we ensure that we capture the majority of these temperate planets while excluding the high-irradiation tail of the distributions.} We also adopt an upper equilibrium temperature boundary of 400~K, and a lower instellation boundary of $S_{\rm p}=0.1 ~\rm S_{\oplus}$ to incorporate the extremes of the TRAPPIST-1 planets. 

\textcolor{black}{This definition is both statistical and astrophysical. The upper $T_{\rm eq}$ limit follows \cite{gunther2019}, who motivated it by the 395~K survivability limit of terrestrial extremophiles. Additionally, all seven TRAPPIST-1 planets fall in this regime, where 3D climate models predict that liquid water could exist on tidally locked worlds through dayside cloud feedback \citep{Yang2013} or terminator habitability \citep{Menou2013,Lobo2023}. Thus our `temperate zone' definition extends the classical habitable zone concept beyond the limits defined by \cite{kopparapu2014} to incorporate a wider diversity of potentially clement worlds.}

\section{TEMPOS: Temperate M dwarf Planets with SPECULOOS} \label{sec:tempos}

\color{black}

As described briefly in Section~\ref{sec:intro}, this paper aims to introduce for the first time a programme running on the SPECULOOS telescopes, TEMPOS (\textbf{TE}mperate \textbf{M} dwarf \textbf{P}lanets with SPECULO\textbf{OS}). 
This programme aims to produce a catalogue of precise temperate exoplanet radii for planets transiting mid-to late- type M dwarfs. Ultimately, we aim to achieve precisions $\lesssim3\%$ \citep[\textit{PLATO} goal,][]{plato2025}, but have a minimum requirement of at least $<7\%$. 

All targets in our TEMPOS sample are obtained from the \textit{TESS} Objects of Interest (TOIs) list. We retain TOIs that meet the following criteria: $T_\mathrm{eff}\leq3400\,\rm K$, $0.1\leq S/\mathrm{S}_\oplus\leq5$, and are confirmed planets or planet candidates.
This sample is not planet-radius limited. At the time of writing, the TEMPOS sample contains 40 confirmed or candidate planets (21 and 19 respectively). 
For the confirmed planets, we update the TOI parameters with published values.

From the available planet radii in this sample, we find current mean and median precisions of $21.3\%$ and $7.7\%$ respectively. Fewer than half of these planets (19) have precisions less than our minimum target threshold of $7\%$, 13 of which are confirmed planets. 
In total, only 2 planets (both confirmed) have precisions $\leq3\%$.

\textcolor{black}{Why require a planetary radius precision of $\leq7\%$, with a goal of $\leq3\%$?} Many studies require precisely measured exoplanet radii \citep[see e.g.,][]{parc2024}, especially since exoplanet masses are typically much less precise (average precision > $20\%$\footnote{NASA exoplanet archive, Jul 2025.}). A radius precision below $7\%$ is sufficient for exoplanet population-level studies and for selecting targets for atmospheric follow-up with \textit{JWST}.

For smaller planets ($\leq2\,\mathrm{R}_\oplus$), however, a higher precision is crucial. A $\sim3\%$ uncertainty is typically needed to distinguish between purely rocky planets and those with volatile envelopes. For example, \citet{rogers2015} show that a $R_\mathrm{p}=1.6\,\mathrm{R}_\oplus$ planet with a $5\%$ uncertainty could sit either side of the radius valley, at either $R_\mathrm{p}=1.5\,\mathrm{R}_\oplus$ or $R_\mathrm{p}=1.7\,\mathrm{R}_\oplus$. \textcolor{black}{The radius valley is defined as the observed dearth in close-in ($P<100\,$days) exoplanets with radii between $1.5<R_\mathrm{p}<2\,\mathrm{R}_\oplus$ \citep{Fulton2017}, and is thought to separate terrestrial super-Earths and gaseous sub-Neptunes, potentially through mechanisms such as photoevaporation of primordial H/He atmospheres  \citep[e.g.,][]{chen2016,Owen2017}, core-powered mass loss \citep{gupta2019} or ``gas-poor" formation \citep{lopez2018}. Thus, a planet with a $5\%$ radius uncertainty could be classified as either a super-Earth or sub-Neptune.} Therefore, in order accurately distinguish between these compositions, we adopt a target precision of $3\%$ on planet radius, which matches the \textit{PLATO} mission goals. While a $\sim7\%$ precision does not necessarily allow for finer differentiation between planet compositions, it is sufficient for robust statistical studies of small planet demographics.

\color{black}

\section{Stellar characterisation} \label{sec:stellar_char}
\textcolor{black}{TOI-6716 (TIC 112115898) and TOI-7384 (TIC 192833836) are two M dwarfs of spectral type M4, located at $\rm 18.9\,pc$ and $\rm 66.8\,pc$ from the Sun respectively \citep{BJdist}. As all our planetary information will be derived using the host stars' parameters, we begin in the sections that follow by characterising TOI-6716 and TOI-7384. All photometric and stellar parameters adopted for this work can be found in Table \ref{tab:starpar}.}

\begin{table*}
\centering
\caption{Stellar parameters adopted for this work. 
}
\begin{tabular}{@{}lp{25mm}p{30mm}p{30mm}@{}}
\toprule
{\bf Star} & {\bf TOI-6716} & {\bf TOI-7384} & \\
\toprule
{\bf Designations} & \multicolumn{1}{p{50mm}}{TIC 112115898, 2MASS J07263809-3033087, Gaia DR2 5605438925569442432, UCAC4 298-022880} & \multicolumn{1}{p{50mm}}{TIC 192833836, 2MASS J05322483-3950016, Gaia DR2 4808732662634331520, UCAC4 251-006761, WISE J053224.78-395002.7} & \\ \midrule
{\bf Parameter} & {\bf Value}    &  {\bf Value}  & {\bf Source} \\ \midrule
T mag      &  11.774$\pm$0.007   & 13.694$\pm$0.008        & \cite{TICv8} \\
B mag      &  16.362$\pm$0.061   & 17.63$\pm$0.17      & \cite{ucac4} \\
V mag      &  14.8$\pm$0.2   & 16.704$\pm$0.206        & \cite{ucac4} \\
G mag      &  13.1269$\pm$0.0006   & 15.0456$\pm$0.0008      & \cite{gaiaDR3cat} \\
J mag      &   10.093$\pm$0.026  & 12.001$\pm$0.026          & \cite{2masscat} \\
H mag      &  9.535$\pm$0.026   & 11.382$\pm$0.026          & \cite{2masscat} \\
K mag      &  9.196$\pm$0.021   & 11.096$\pm$0.023         & \cite{2masscat} \\
W1 mag     &   -   & 10.942$\pm$0.023          & \cite{wisecat} \\
W2 mag     &    -  & 10.779$\pm$0.021          & \cite{wisecat} \\
W3 mag     &    -  & 10.551$\pm$0.064            & \cite{wisecat} \\
W4 mag     &    -  & > 8.92       & \cite{wisecat} \\
Distance   &  18.89$\pm$0.02\,pc    & 66.79$\pm$0.20\,pc       & \cite{BJdist} \\
$\alpha$    &    07:26:37.7    & 05:32:24.76      & \cite{gaiaDR3cat} \\
$\delta$     &  -30:33:07.85     & -39:50:03.3      & \cite{gaiaDR3cat} \\
$\mu_{\alpha}$   &    	$\rm -316.5\,mas\,yr^{-1}$    & $\rm -46.0\,mas\,yr^{-1}$      & \cite{gaiaDR3cat} \\
$\mu_{\delta}$   &   $\rm 59.0\,mas\,yr^{-1}$     & $\rm -101.8\,mas\,yr^{-1}$      & \cite{gaiaDR3cat} \\
SpT      &  M4     & M4                  & This work (opt. spec.) \\
     &  M4     & M3$\pm0.5$                  & This work (NIR spec.) \\
$R_{\star}$  & $0.231\pm0.015\,\mathrm{R}_{\odot}$  & $0.319\pm0.018\,\mathrm{R}_{\odot}$ & This work (SED)             \\

$M_{\star}$  & $0.223\pm0.0110\,\mathrm{M_{\odot}}$  & $0.318\pm0.016\,\mathrm{M}_{\odot}$   & This work (SED)   \\

${\rm T_{eff}}$ & 3110$\pm$80\,K & 3185$\pm$75\,K             & This work (SED)              \\
 
$\log g_\star$   &   5.06$\pm$0.06     & 4.93$\pm$0.05         & This work (from SED)                 \\

$\rm [Fe/H]$     & 0.0$\pm$0.5 dex  & 0.0$\pm$0.5 dex              & This work (SED)                 \\
                & +0.17$\pm$0.20 dex  & +0.17$\pm$0.20 dex              & This work (opt. spec.)                 \\
                & -0.09$\pm$0.12 dex  & +0.18$\pm$0.11 dex              & This work (NIR spec.)                 \\

\bottomrule
\end{tabular}
\label{tab:starpar}
\end{table*}

\color{black}
\subsection{Reconnaissance spectroscopy}

\subsubsection{TOI-6716}

\begin{figure}
    \centering
    \includegraphics[width=\linewidth]{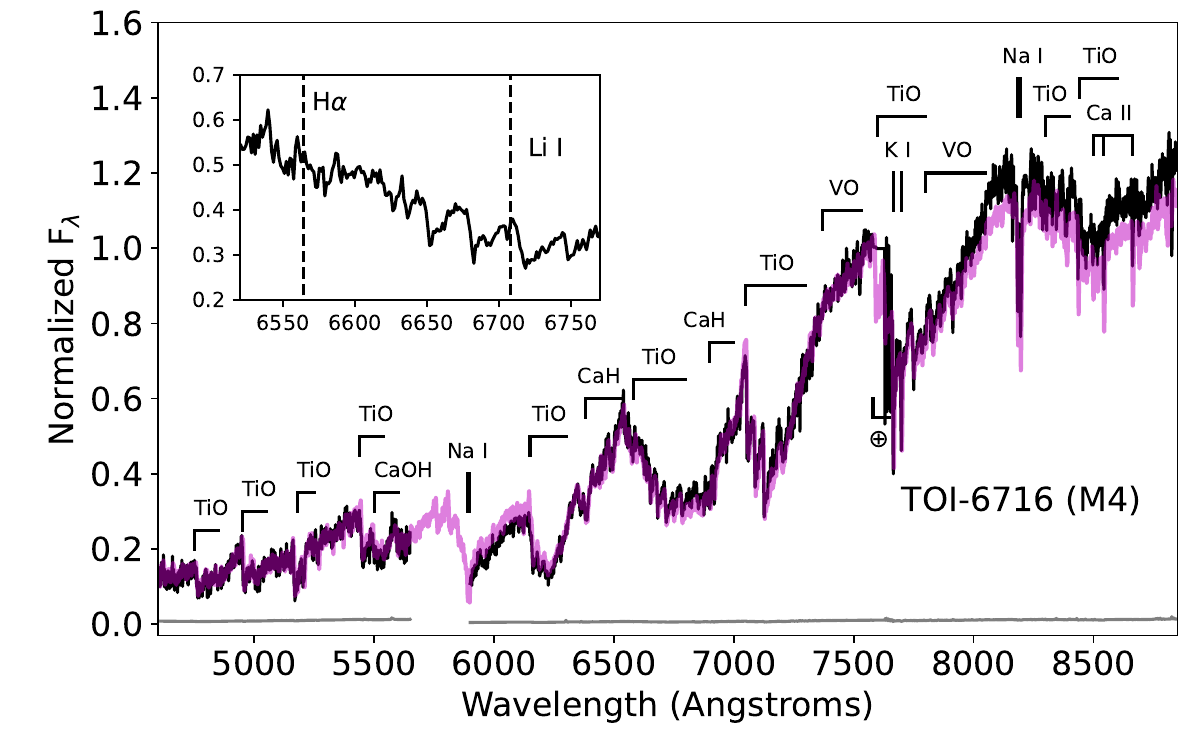}
    \includegraphics[width=\linewidth]{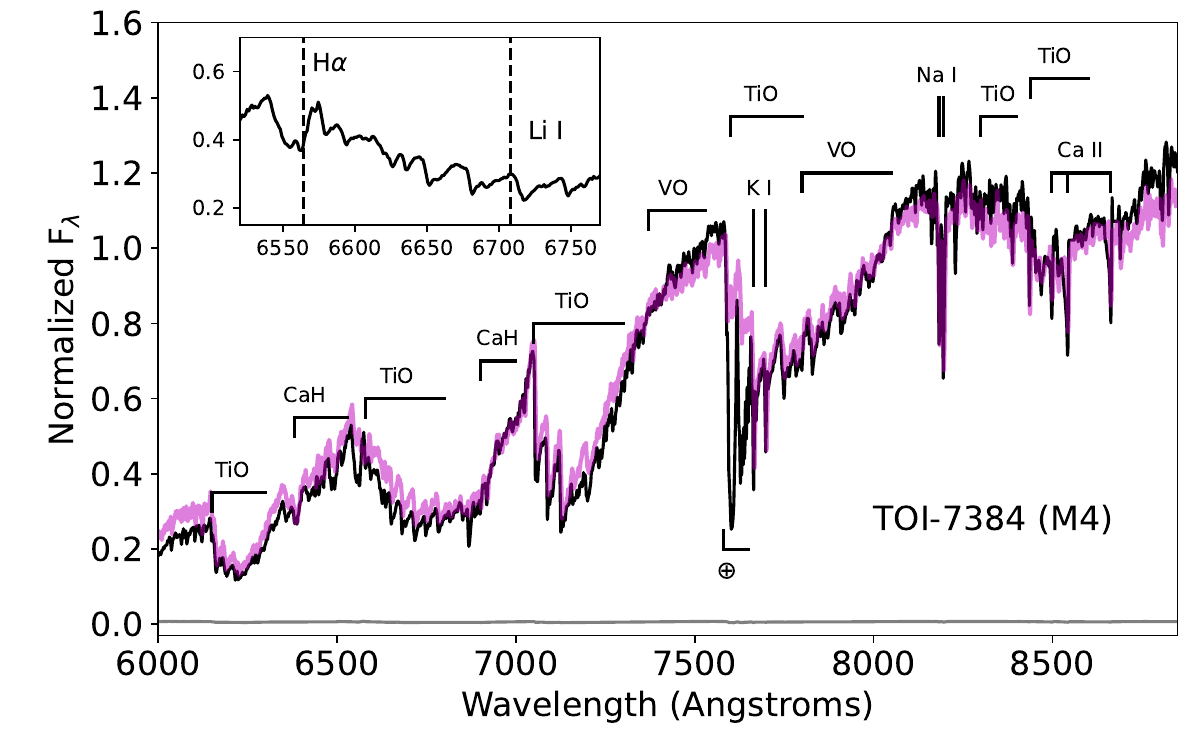}
\caption{Shane/Kast optical spectrum of TOI-6716 (top) and Magellan/LDSS-3 spectrum of TOI-7384 (bottom), compared to their best-fit M4 SDSS spectral template from \citet[magenta line]{2007AJ....133..531B}. Key spectral features are labelled, including regions of residual telluric absorption ($\oplus$). 
Inset boxes show the 6520--6770~{\AA} region encompassing H$\alpha$ and Li~I features.
The gap in the Kast spectrum  between 5600~{\AA} and 5900~{\AA} corresponds to the gap between that instrument's blue and red channels.
}
    \label{fig:kast-ldss3}
\end{figure}

We observed TOI-6716 with the Kast double spectrograph \citep{kastspectrograph} on the 3-m Shane telescope at Lick Observatory on 2025 Mar 08 (UT) in clear conditions with 1.2$\arcsec$ seeing. We used the 1.5$\arcsec$ slit aligned to the parallactic angle to obtain blue and red optical spectra split at 5700\,{\AA} by the d57 dichroic, and dispersed by the 600/4310 grism and 600/7500 grating, resulting in spectral resolutions of $\lambda/\Delta\lambda$ $\approx1100$ and $\approx1500$ for the blue and red spectra, respectively. We obtained a single 500\,s exposure in the blue channel and two 250\,s exposures in the red channel at an average airmass of 2.7. The G2\,V star HD 60513 ($V=6.7$) was observed at a slightly lower airmass of 2.0 for telluric absorption calibration, and the spectrophotometric calibrator Hiltner~600 \citep{1992PASP..104..533H,1994PASP..106..566H} was observed shortly thereafter for flux calibration. We used HeHgCd and HeNeArHg arc lamp exposures to wavelength calibrate our blue and red data, and flat-field lamp exposures for pixel response calibration. Data were reduced using the \texttt{kastredux} code\footnote{\url{https://github.com/aburgasser/kastredux}.} using standard settings. The resulting spectra have median signal-to-noise ratios (SNRs) of
22 at 5425\,{\AA} and 88 at 7350\,{\AA}.

The reduced spectrum is shown in Fig.\,\ref{fig:kast-ldss3}, with a comparison to the best-fit M4 dwarf SDSS spectral template from 
\citet{2007AJ....133..531B}.
This classification was confirmed using index-based methods described in 
\citet{1995AJ....110.1838R,1997AJ....113..806G,1999AJ....118.2466M,2003AJ....125.1598L}; and \citet{2007MNRAS.381.1067R}, which span M3 to M4.
We detect a weak signature of H$\alpha$ emission at 6563~{\AA} with an equivalent width EW = $-1.20\pm0.18$~{\AA}, corresponding to $\log{\left(L_{{\rm H}\alpha}/L_{\rm bol}\right)} = -4.37\pm0.09$ using the $\chi$ factor relation of \citet{2014ApJ...795..161D}. 
The presence of weak H$\alpha$ emission indicates an activity age of no more than 4--6~Gyr \citep{2008AJ....135..785W}, while the absence of detectable Li\,\textsc{i} absorption at 6708~{\AA} rules out a substellar mass and age less than $\sim30$\,Myr. We measure the metallicity index $\zeta=1.124\pm0.005$  \citep{2013AJ....145..102L}, which corresponds to a roughly solar metallicity of [Fe/H]$=+0.17\pm0.20$ using the \citet{Mann2013} calibration, consistent with solar metallicity. \\

\begin{figure}
    \centering
    \includegraphics[width=\linewidth]{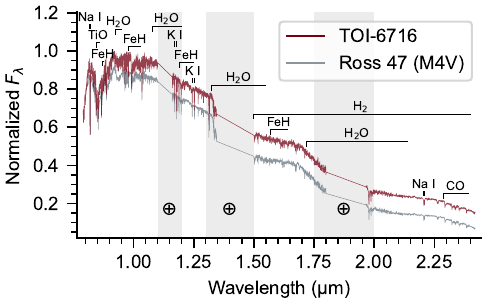}
    \includegraphics[width =\columnwidth]{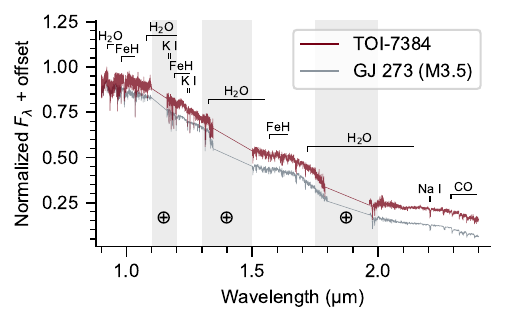}
    \caption{SpeX/SXD spectrum of TOI-6716 (top) and Magellan/FIRE spectrum of TOI-7384 (bottom). The target spectra (red) are compared to M4V standard Ross 47 and the M3.5 standard GJ\,273, respectively, both shown in grey and offset vertically. Strong M-dwarf spectral features and spectral regions with strong telluric absorption are indicated.}
    \label{fig:spex-fire}
\end{figure}

We also observed TOI-6716 with the SpeX spectrograph \citep{Rayner2003} on the 3.2-m NASA Infrared Telescope Facility (IRTF) on 2024 Nov 10 (UT) under clear conditions with $0.5''$ seeing.
Using the short-wavelength cross-dispersed (SXD) mode and the $0\farcs3 \times 15''$ slit ($\lambda/\Delta\lambda{\sim}2000$, 0.80--2.42\,$\mu$m) aligned to the parallactic angle, we obtained eight 60-s exposures at an airmass of 1.6, nodding in an ABBA pattern.
We followed with a standard set of SXD flat field and arc lamp calibrations and twelve 30-s exposures of the A0\,V standard HD\,56751 ($V{=}7.1$) at a similar airmass.
Data reduction with the Spextool v4.1 pipeline \citep{Cushing2004} followed the standard approach \citep{Barkaoui2024, Barkaoui2025, Ghachoui2024}.
The final spectrum has a median SNR per pixel of 144.

The SXD spectrum of TOI-6716 is shown in Fig.\,\ref{fig:spex-fire}.
With respect to single-star standards in the IRTF Spectral Library \citep{Cushing2005, Rayner2009}, we find the closest match to the M4\,V standard Ross\,47 using the SpeX Prism Library Analysis Toolkit \citep[SPLAT, ][]{splat} and adopt a spectral type of M4.0\,$\pm$\,1.0, consistent with the optical classification.
From the $K$-band Na\,\textsc{i} and Ca\,\textsc{i} lines and the H$_2$O--K2 index \citep{Rojas-Ayala2012}, we estimate $\mathrm{[Fe/H]} = -0.09 \pm 0.12$ using the \citet{Mann2013} relation and the Monte Carlo approach detailed previously \citep{Delrez2022, Ghachoui2023}, again consistent with solar metallicity.

\subsubsection{TOI-7384}

We gathered an optical spectrum of TOI-7384 using the Low Dispersion Survey Spectrograph \citep[LDSS-3C,][]{Stevenson2016} on the 6.5-m \textit{Magellan II} (Clay) Telescope on 2022 Jan 07 (UT) during clear conditions with $0\farcs5$ seeing.
We used the standard setup for the long-slit mode (fast readout speed, low gain, and $1{\times}1$ binning) along with the VPH-Red grism, OG-590 blocking filter, and the $0\farcs75 \times 0\farcm4$ center slit.
This configuration provides spectra covering 6000--10\,000\,\AA{} with a spectral resolving power of $R{\sim}1810$.
We collected six, 300-s exposures at an average airmass of 1.023.
We followed the science observations with three, 1-s exposures of the F8\,V standard HR\,1651 \citep{Gray2006}.
At each pointing, we collected a 1-s HeNeAr arc lamp exposure and three, 10-s flats with the ``quartz high'' lamp.
We used a custom, Python-based pipeline \citep{Dransfield2023} to reduce the data, including bias removal, flat-field correction, and spectral extraction.
We used the HeNeAr arc exposure for wavelength calibration.
While we did not apply a telluric correction, we used the ratio of the F8\,V spectrum to the F8\,V template from \citet{Pickles1998} to calculate a relative flux correction.
The final spectrum has a maximum SNR per pixel of 177 at 9183\,\AA{}, a mean SNR per pixel of 113 in the 6000--10\,000\,\AA{} range, and an average of 2.3 pixels per resolution element.

The LDSS-3 spectrum of TOI-7384 is also shown in Fig.\,\ref{fig:kast-ldss3}, also compared to its best-fit M4 dwarf SDSS spectral template, which is again consistent with index-based classifications spanning M3.5--M4.5. We detect H$\alpha$ in absorption in this spectrum (EW = +1.12$\pm$0.17~{\AA}) indicating weak or absent magnetic activity and an activity age of at least 4.5~Gyr \citep{2008AJ....135..785W}. For this source we measure $\zeta=1.127\pm0.003$, which again corresponds to a roughly solar metallicity of [Fe/H]$=+0.17\pm0.20$ using the \citet{Mann2013} calibration. \\

We also observed TOI-7384 with the FIRE spectrograph \citep{Simcoe2008} on the 6.5-m Magellan Baade Telescope on 2023 Feb 02 (UT) under clear conditions with 1$\farcs$0 seeing.
We used the high-resolution echellette mode with the 0$\farcs$45 slit, providing 0.9--2.4\,$\mu$m spectra with a resolving power of $R{\sim}8000$.
We collected two exposures of 306.5-s each, nodding between frames.
Following the science observations, we collected a 30-s arc lamp exposure and two, dithered exposures of the A0\,V standard HD\,38433 ($V{=}9.4$).
Along with a set of internal and dome flats collected the previous afternoon, we used these calibrations to reduce the data with the \texttt{FIREHOSE} pipeline\footnote{\url{https://github.com/rasimcoe/FIREHOSE}}.
The final spectrum has a median SNR per pixel of 151 and an average of 2.4\,pixels per resolution element.

The FIRE spectrum of TOI-7384 is shown in Fig.\,\ref{fig:spex-fire}.
We used SPLAT to compare the spectrum to SXD spectra of single-star standards in the IRTF Spectral Library \citep{Cushing2005, Rayner2009}.
We found the best spectral match to the M3.5 standard GJ\,273 with similar but poorer matches to the M3 standard GJ\,388 and M4 standard GJ\,213, and so we adopt an infrared spectral type of M3.5 $\pm$ 0.5, again consistent with the optical type.
To address an issue with continuum normalization at the long-wavelength end of the spectrum, we applied a correction to the data at wavelengths ${>}2.25\,\micron$ using the spectrum of GJ\,273 as a reference.
We used the corrected spectrum to estimate the stellar metallicity using the \citet{Mann2013} relation between the equivalent widths of the $K$-band Na\,\textsc{i} and Ca\,\textsc{i} doublets and the H2O--K2 index \citep{Rojas-Ayala2012}.
This gives an iron abundance of $\mathrm{[Fe/H]} = +0.18 \pm 0.11$, consistent with the optical metallicity.

\color{black}
\subsection{Spectral Energy Distribution}
\label{sec:sed}

\begin{figure}
    \centering
    \includegraphics[width=\columnwidth,trim=80 70 50 50,clip
    ]{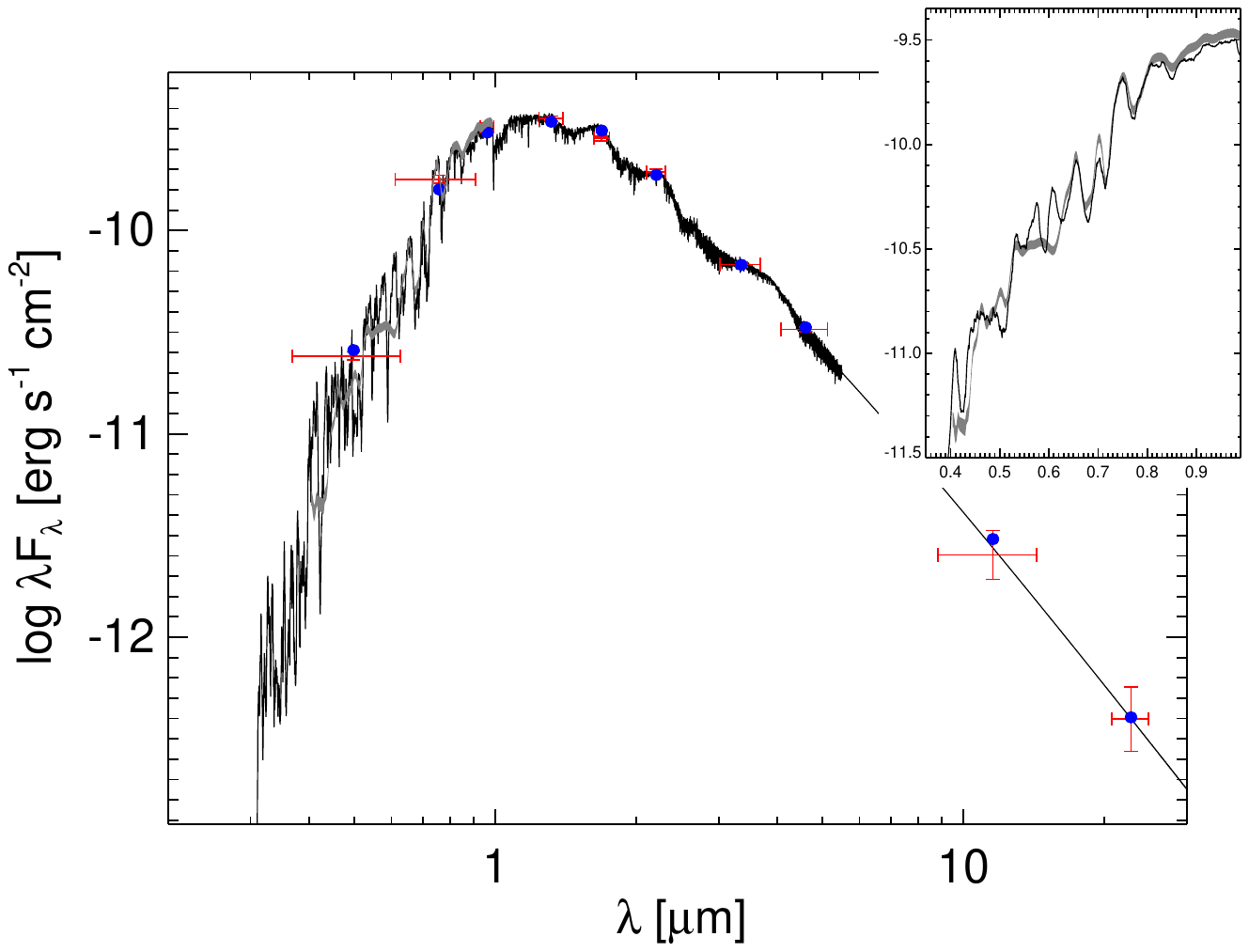}

\includegraphics[width=\columnwidth,
trim=80 70 50 50,clip
]{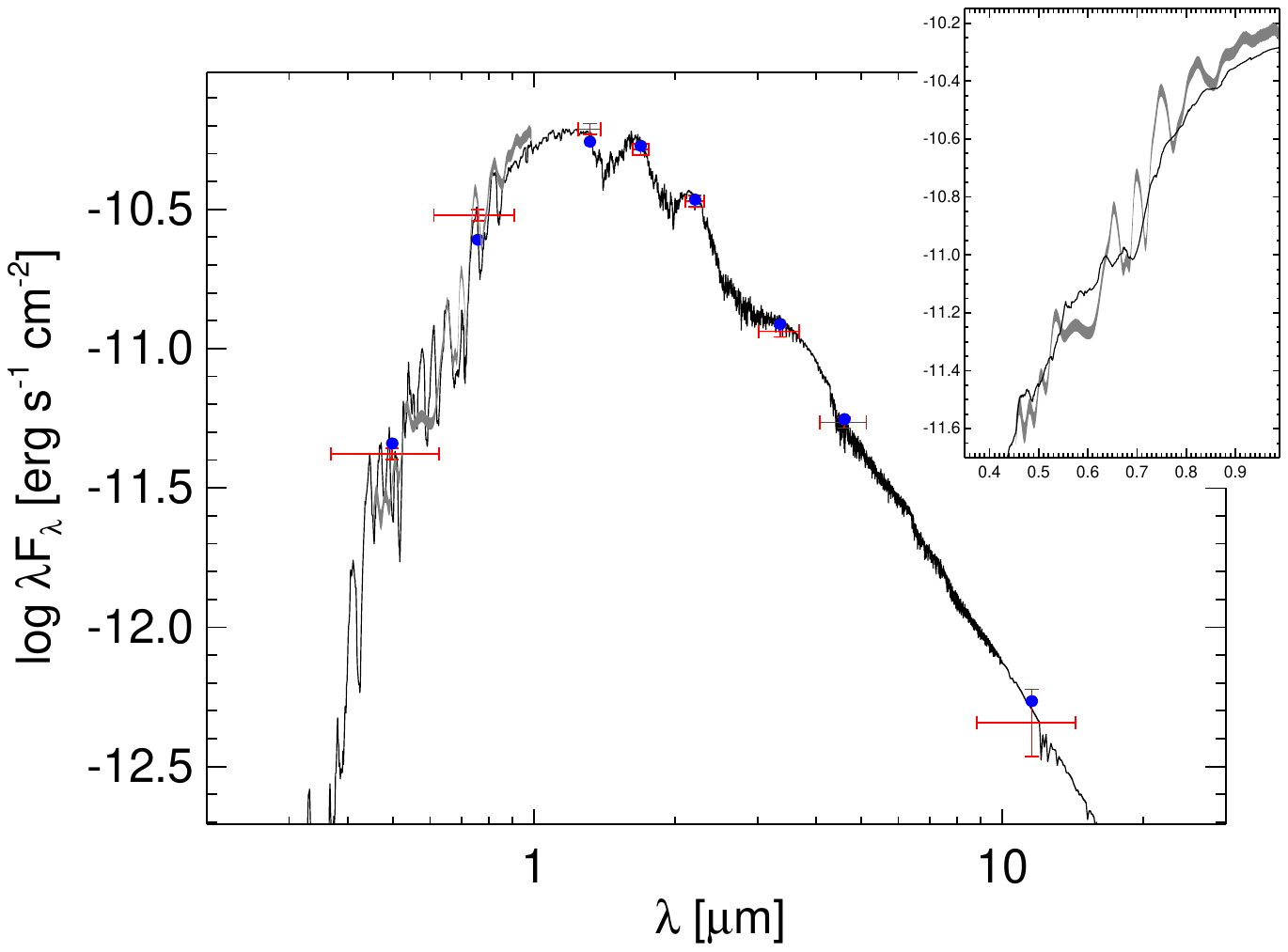}
\caption{Spectral energy distributions of TOI-6716 (top) and TOI-7384 (bottom). Red symbols represent the observed photometric measurements, where the horizontal bars represent the effective width of the passband. Blue symbols are the model fluxes from the best-fit NextGen atmosphere model (black). The inset axes show the absolute flux-calibrated {\it Gaia\/} spectrophotometry as a grey swathe overlaid on the best-fit model. \label{fig:sed}}
\end{figure}

As an independent determination of the basic stellar parameters, we performed an analysis of the broadband spectral energy distribution (SED) of each of the stars together with the {\it Gaia\/} DR3 parallax \citep[with no systematic offset applied; see, e.g.,][]{StassunTorres:2021}. This yielded, upon following the procedures described in \citet{Stassun:2016,Stassun:2017,Stassun:2018}, an emperical measurement of each stellar radius.
We pulled the the $JHK_S$ magnitudes from {\it 2MASS}, the W1--W3 magnitudes from {\it WISE}, and the $G_{\rm BP} G_{\rm RP}$ magnitudes from {\it Gaia}. Finally, we utilized the absolute flux-calibrated {\it Gaia\/} spectrophotometry where available. Together, the available photometry spans the full stellar SED over the wavelength range 0.4--10~$\mu$m (see Fig.~\ref{fig:sed}). 

We performed a fit using NextGen stellar atmosphere models \citep{HauschildtAllardBaron}, with the free parameters being the effective temperature ($T_{\rm eff}$) and metallicity ([Fe/H]), as well as the extinction $A_V$, which we limited to maximum line-of-sight value from the Galactic dust maps of \citet{Schlegel:1998}. Integrating the (unreddened) model SED gives the bolometric flux at Earth, $F_{\rm bol}$. Taking the $F_{\rm bol}$ and $T_{\rm eff}$ together with the {\it Gaia\/} parallax, gives the stellar radius, $R_\star$. In addition, we can estimate the stellar mass from the empirical $M_K$ relations of \citet{Mann2019}. 
\textcolor{black}{For TOI-6716\,b,} the resulting fit (Fig.~\ref{fig:sed}) has a best-fit $A_V \equiv 0$, $T_{\rm eff} = 3110 \pm 80$~K, [Fe/H] = $0.0 \pm 0.5$, with a reduced $\chi^2$ of 2.2. Thus, we find $F_{\rm bol} = 4.027 \pm 0.094 \times 10^{-10}$ erg~s$^{-1}$~cm$^{-2}$, $R_\star = 0.231 \pm 0.015$~R$_\odot$, and $M_\star = 0.223 \pm 0.011$~M$_\odot$. 
\textcolor{black}{For TOI-7384\,b,} the resulting fit (Fig.~\ref{fig:sed}) has a best-fit $A_V = 0.01 \pm 0.01$, $T_{\rm eff} = 3185 \pm 75$~K, [Fe/H] = $0.0 \pm 0.5$, with a reduced $\chi^2$ of 1.9. Thus, we find $F_{\rm bol} = 6.75 \pm 0.40 \times 10^{-11}$ erg~s$^{-1}$~cm$^{-2}$, $R_\star = 0.319 \pm 0.018$~R$_\odot$, and $M_\star = 0.318 \pm 0.016$~M$_\odot$. 

\textcolor{black}{While a set of stellar parameters may also be obtained using empirical relations, we choose to use those calculated from the SED. This choice is motivated by the fact that empirical relations rely on the GAIA DR3 \textit{BP-RP} colour to estimate the $T_\mathrm{eff}$, which is known to be unreliable for cooler stars \citep[see e.g.,][]{jordi2010}, and is highly sensitive to metallicity, which is challenging to robustly derive for M dwarfs \citep[see e.g.,][]{lindgren_heiter2017, passegger2022}. In contrast, the SED fitting approach uses a wide range of broadband photometry and allows both $T_\mathrm{eff}$ and metallicity to be free parameters. The resulting stellar parameters derived from this are thus more robust and have more conservative uncertainties.}

\section{Planet identification} 
\label{sec:planet_identification}

\subsection{Candidate identification} \label{tess}
\color{black}

Prior to TOI-7384\,b being identified as a planet candidate by \textit{TESS}, observations were first triggered by the NEMESIS (Exoplanet Tra\textbf{N}sit Survey of N\textbf{E}arby \textbf{M} Dwarfs in \textit{T\textbf{ES}S} FF\textbf{Is}) pipeline, presented in \cite{nemesis_pipeline}. This pipeline was designed to produce detrended photometry of M dwarfs visible in \textit{TESS} full frame images (FFIs) and conduct a transit search on the resulting lightcurves. In the first release the pipeline analysed \textit{TESS} data from Sectors 1--5, which included 33,054 M dwarfs within $100\,\rm pc$, and produced 29 planet candidates. Of these, 5 matched known \textit{TESS} Objects of Interest (TOIs) and the remaining 24 were new detections. We selected the 12 candidates found orbiting the hosts with effective temperatures $<3300\,\rm K$  as these aligned the best with the existing SPECULOOS target list. These were given internal designations of NEMESIS 1--12 for the purpose of our follow-up campaign. Nemesis-12 was identified as a planet candidate and designated the CTOI (Community \textit{TESS} Object of Interest) TIC 192833836.01. In April 2025, it was identified as TOI-7384.01, which will remain its designation throughout this paper.

\begin{figure*}
    \centering
    \includegraphics[width=\textwidth]{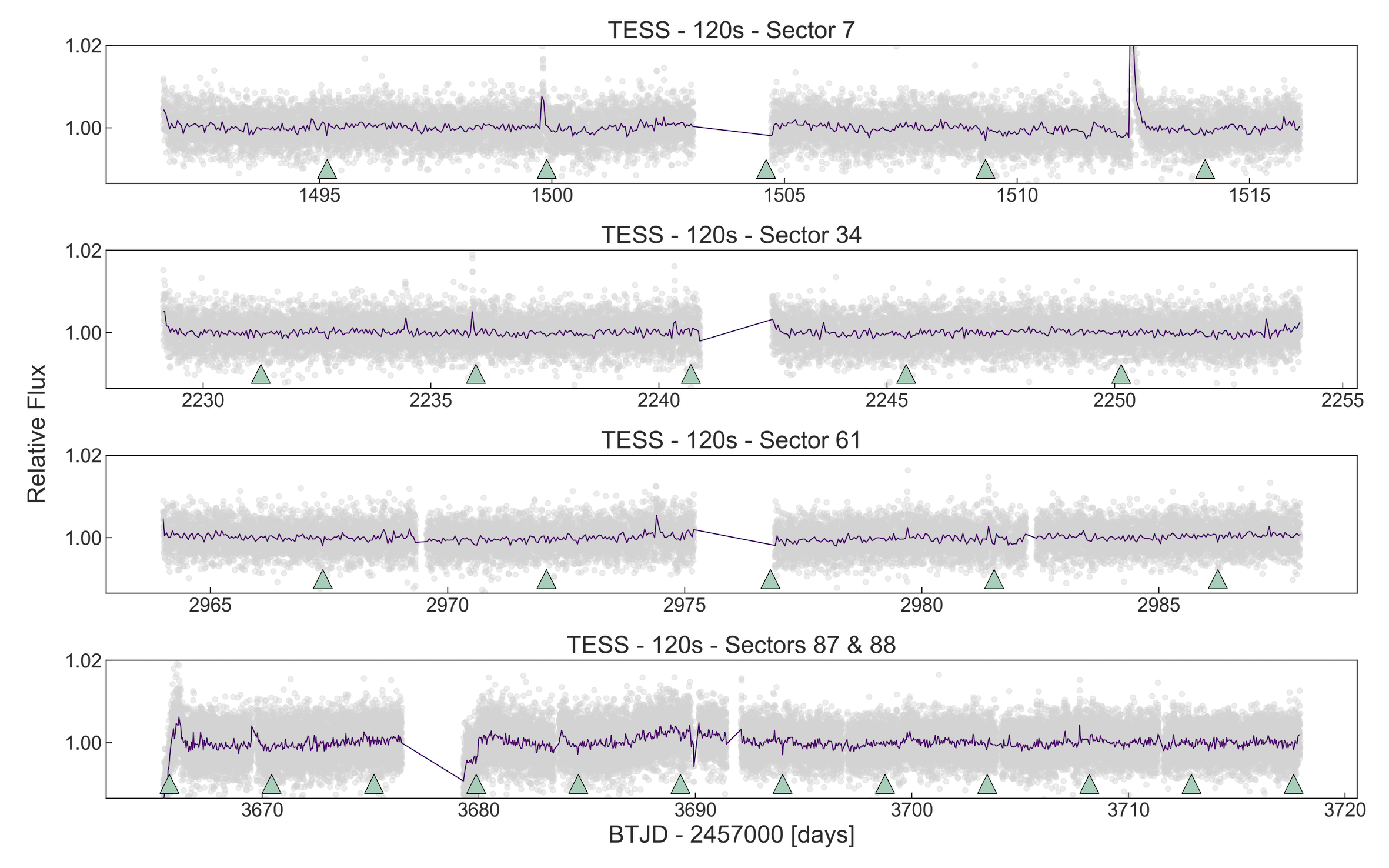}
    \caption{TESS 2-minute cadence photometry (grey) for TOI-6716 for Sectors 7 (top), 34 (top-middle), 61 (bottom-middle), 87 and 88 (bottom). Purple shows the data binned by one hour. Transits are indicated with green triangles, although are not visually obvious in the \textit{TESS} data.}
    \label{fig:tess_6716}
\end{figure*}

\begin{figure*}
    \centering
    \includegraphics[width=\textwidth]{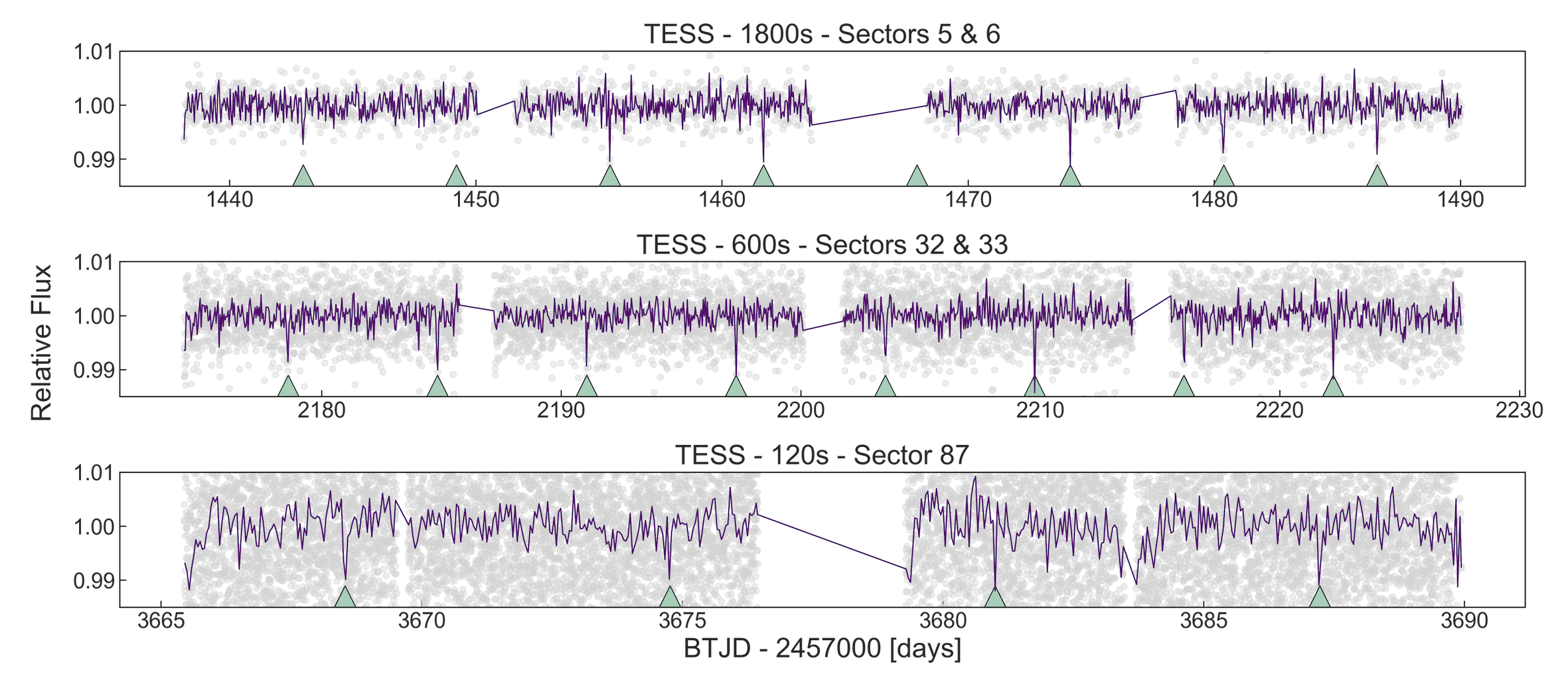}
    \caption{TESS photometry (grey) for TOI-7384 in 2-minute cadence (top), 10-minute cadence (middle) and 30-minute cadence (bottom). Purple shows the data binned by one hour. Transits are indicated with green triangles.}
    \label{fig:tess_7384}
\end{figure*}

TOI-7384 was observed in \textit{TESS} full-frame images in Sectors 5 (as mentioned above) and 6 in 1800s cadence, 32 and 33 in 600s cadence, and 87 in 200s and 120s cadence. In Fig.~\ref{fig:tess_7384} we present the \textit{TESS} photometry as reduced by the TESS-SPOC (Sectors 5,6,32,33) and SPOC (Sector 87) pipelines. TOI-7384\,b has a candidate orbital period and planetary radius of $P_{b,\,7384}=6.23\,$d and $R_{b,\,7384}=3.93\,\mathrm{R}_\oplus$ respectively. 
TOI-6716 was observed in \textit{TESS} full-frame images at cadence 1800s in Sector 7, 600s in Sector 34, 200s in Sectors 61, 87 and 88, 120s in Sectors 7, 34, 61, 87 and 88, and 20s in Sectors 87 and 88. It was reported as a planet candidate on 2023 Oct 5. Fig. ~\ref{fig:tess_6716} shows the \textit{TESS} 120s cadence photometric observations reduced by the SPOC (Science Processing Operations Center) pipeline \citep{SPOC}. It has a candidate orbital period and planetary radius of $P_{b,\,6716}=4.72\,$d and $R_{b,\,6716}=1.01\,\mathrm{R}_\oplus$ respectively. 
The data for both TOI-6716 and TOI-7384 are available on the NASA Mikulski Archive for Space Telescopes.

\begin{figure*} 
  \centering
  \subfloat[Target pixel file image for TOI-6716.]{\includegraphics[width=\columnwidth]{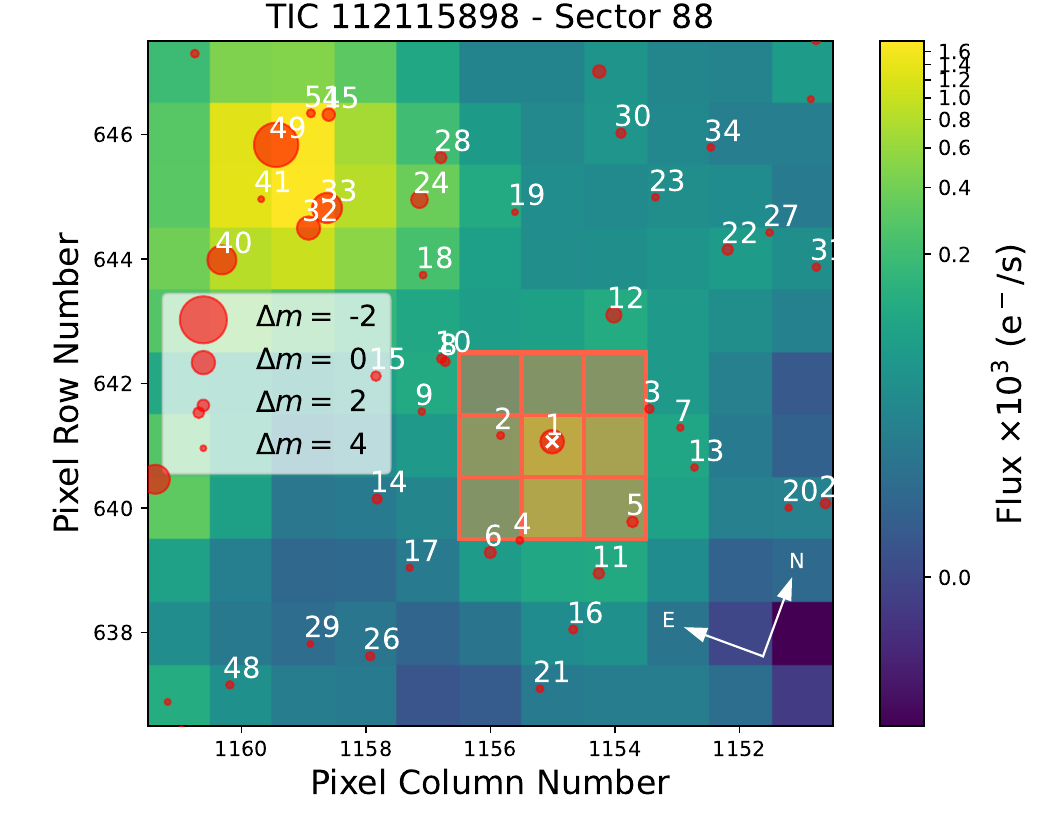}}\label{fig:6716_tpf}\quad 
  \subfloat[Target pixel file image for TOI-7384.]{\includegraphics[width=\columnwidth]{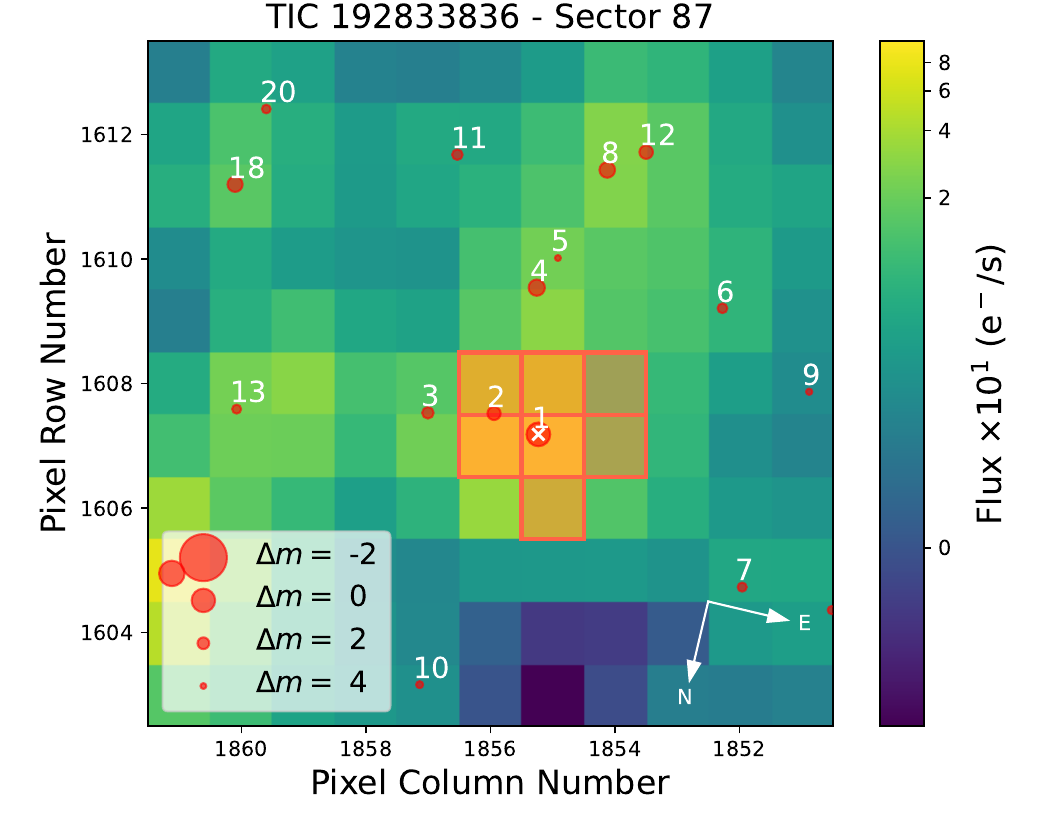}} \label{fig:7384_tpf}
  \caption{\textit{TESS} target pixel file image created with \textsc{tpfplotter} for TOI-6716 (\textit{left}) and TOI-7384 (\textit{right}) observed in Sector 88 (2025 Jan 14 -- 2025 Feb 11) and Sector 87 (2024 Dec 18 -- 2025 Jan 14) respectively. The target stars are indicated by a white cross, and the \textit{TESS} aperture is represented by the red hatch. Red dots highlight GAIA-DR3 sources with varying sizes corresponding to their magnitude relative to the respective target stars.}
  \label{fig:tpfplotter}
\end{figure*}

\color{black}
We use the python package \textsc{tpfplotter}\footnote{\textsc{tpfplotter} is publicly available at \url{https://github.com/jlillo/tpfplotter}} \citep{aller2020} to plot the field of view around each star along with the \textit{TESS} apertures from their most recent \textit{TESS} sectors at the time of writing; these are shown in Fig.~{\ref{fig:tpfplotter}}. We can conclude that there is no significant contamination from bright sources in or near the target stars / \textit{TESS} aperture which could be either diluting our transit event, or be the source of the transit event itself. This is discussed further in Section~\ref{sec:validation}.

\subsection{Search for additional candidates and detection limits}

We used the custom pipeline \texttt{SHERLOCK\footnote{\textsc{SHERLOCK} is publicly available at \url{https://github.com/franpoz/SHERLOCK}}} \citep[see, e.g.,][]{sherlockp1,sherlockp2} to analyse the TESS data described in Section~\ref{tess}. This has the dual aims of independently recovering TOI-6716\,b and TOI-7384\,b, and searching for additional planetary candidates potentially overlooked by official pipelines such as SPOC and QLP due to their detection thresholds. The \texttt{SHERLOCK} pipeline is specifically designed to identify low-SNR transit-like signals indicative of planetary transits. It integrates various modules that facilitate data access and rapid inspection, automated transit searches, candidate vetting and validation, Bayesian modeling to derive precise planetary parameters and ephemerides, and the computation of observational windows to guide follow-up campaigns \citep{sherlockp3}.
We successfully recovered the previously known TOI alerts in our initial runs but did not detect any additional transit-like features attributable to planetary origins. 

The lack of additional detections may result from several possibilities \citep[see, e.g.,][]{wells2021,Schanche2021,pozuelos2023}: the system may not contain any additional transiting planets with orbital periods within the range explored here \textcolor{black}{($\leq10\,$days)}, any additional transiting planets, or even any any additional planets.
Alternatively, additional transiting planets may remain undetected due to the limited photometric precision of the data. To assess this possibility, we conducted injection-and-recovery tests with the \texttt{MATRIX} code\footnote{{The \texttt{MATRIX} (\textbf{M}ulti-ph\textbf{A}se \textbf{T}ransits \textbf{R}ecovery from \textbf{I}njected e\textbf{X}oplanets) code is open access on GitHub: \url{https://github.com/PlanetHunters/tkmatrix}}} \citep{devora2022}.

The code explores a three-dimensional parameter space (orbital period, planetary radius, and transit epoch) by generating a grid of synthetic scenarios, which are injected into the original light curve. In our case, the grid comprises 30 periods, 30 radii, and 5 epochs, resulting in 4,500 different scenarios. A synthetic planet is considered as retrieved when a period and an epoch are found that differ by at most 1\% and up to 1 hour from the injected values, respectively.

\begin{figure}
    \centering
    \includegraphics[width=\linewidth]{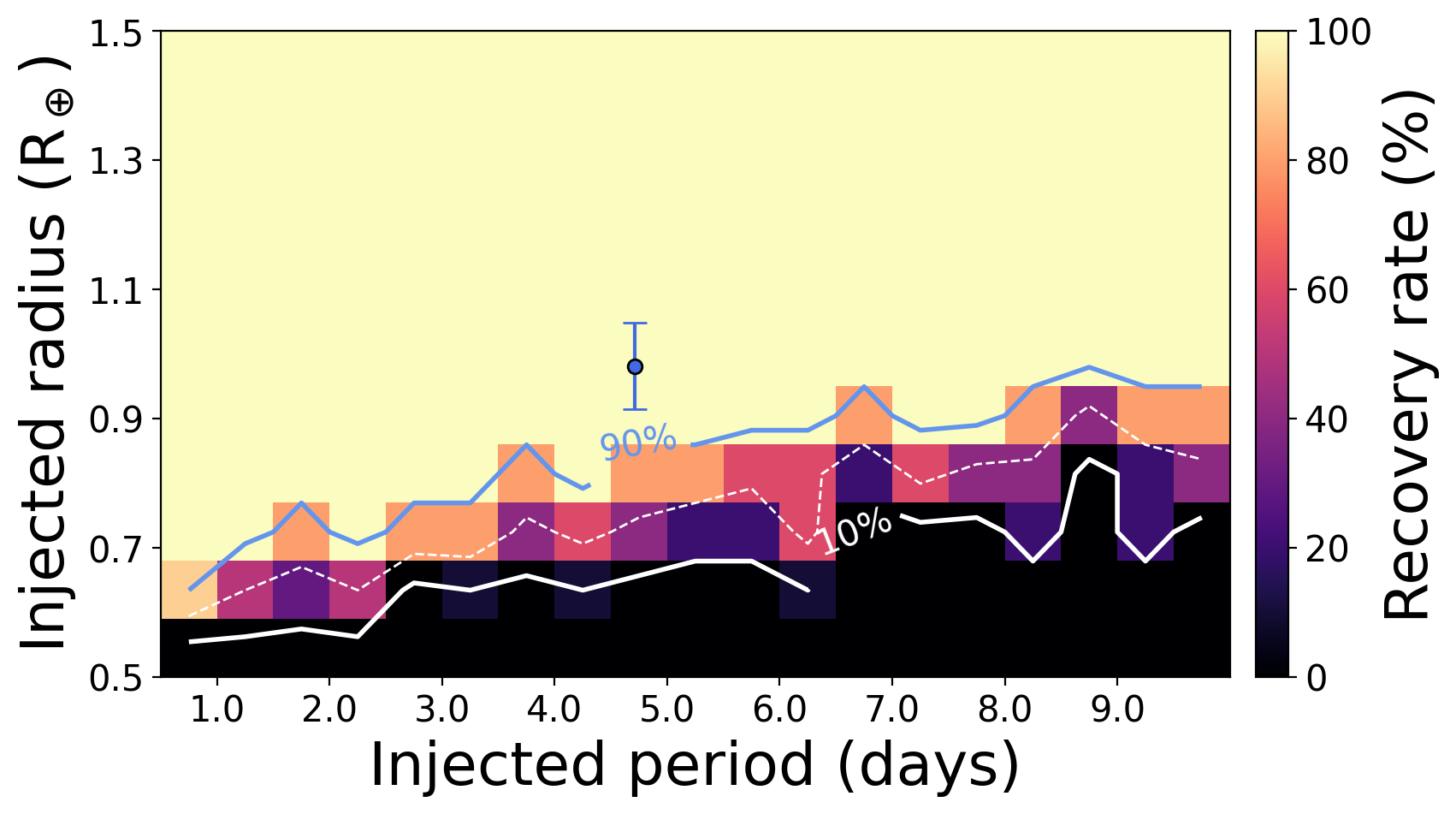}

    \vspace{0.05cm} 
    \includegraphics[width=\linewidth]{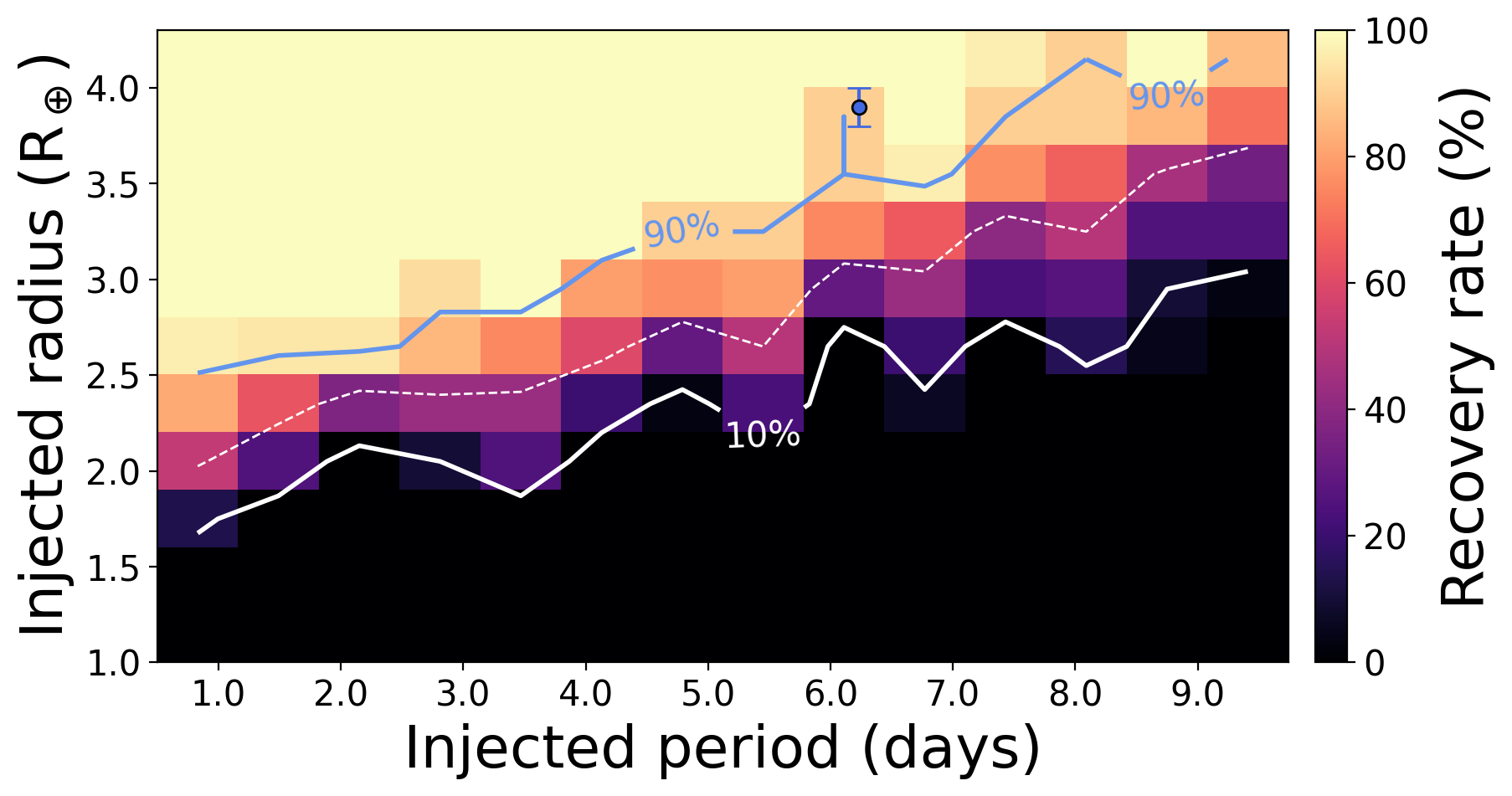}
    \caption{Injection-and-recovery experiment performed to establish the detection limits using all the available \textit{TESS} sectors in each case. Color code indicates recovery rate: bright yellow for high recovery, dark purple/black for low recovery. The solid blue line indicates the 90\% recovery contour, the dashed white line marks the 50\%, and the solid white line corresponds to the 10\%. Blue dots correspond to planets TOI-6716\,b (top) and TOI-7384\,b (bottom).}
    \label{fig:inj_rec}
\end{figure}

The results, displayed in Fig.~\ref{fig:inj_rec}, indicate that for TOI-6716, planets as small as 0.6\,R$_{\oplus}$ are detectable only on short orbits ($P \lesssim 2$\,d), with a 100\% recovery rate for transiting planets larger than 1.0\,R$_{\oplus}$ and $P \leq 10$\,d, effectively ruling out their presence. In contrast, for TOI-7384, the smallest planets recovered with $\geq$90\% rate are $\sim$2.5\,R$_{\oplus}$ at short periods; Earth-sized planets remain undetectable at any period, and their existence cannot be excluded with the current data.

\section{Vetting and validation}
\label{sec:vettingvalidation}

In this section we describe the results of the multi-facility follow-up campaign conducted between October 2023 and April 2025 for TOI-6716\,b, and April 2021 and February 2023 for TOI-7384\,b. We begin with the high-resolution imaging observations, and then outline the photometric observations collected from SPECULOOS Southern Observatory, TRAPPIST-South and LCO. Finally, we describe how all our follow-up observations were used to validate the planetary natures of TOI-6716\,b and TOI-7384\,b.

All follow-up observations are summarised in Tables~\ref{tab:followup-6716} and ~\ref{tab:followup-7384}.

\begin{table*} 
\centering
\caption{Summary of ground-based follow-up observations carried out for TOI-6716.}
\begin{tabular}{@{}ccccc@{}}
\midrule \midrule
\multicolumn{5}{c}{\textbf{TOI-6716 Follow-up Observations}}                                              \\ \midrule \midrule
\multicolumn{5}{c}{\textbf{High Resolution Imaging}}                                             \\ 
\textbf{Observatory} & \textbf{Filter} & \textbf{Date}     & \textbf{Sensitivity Limit} & \textbf{Result}\\ \midrule
Gemini South      & $562~{\rm nm}$     &  & $\Delta m=4.6$ at $0.5\arcsec$  & No sources detected  \\ 
Gemini South      & $832~{\rm nm}$     &  & $\Delta m=7.0$ at $0.5\arcsec$  & No sources detected   \\ \midrule
\multicolumn{5}{c}{\textbf{Photometric Follow-up}}                                               \\
\textbf{Observatory} & \textbf{Filter} & \textbf{Date}     & \textbf{Coverage} & \textbf{Result} \\ \midrule
LCO-CTIO-1m0/SINISTRO & {\it Sloan-$i'$} & 2023 Oct 28 & Full & Detection\\
LCO-SSO-M4/MUSCAT4 & {\it Sloan-$g'$} & 2023 Dec 28 & Full & Detection\\
LCO-SSO-M4/MUSCAT4 & {\it Sloan-$r'$} & 2023 Dec 28 & Full & Detection\\
LCO-SSO-M4/MUSCAT4 & {\it Sloan-$i'$} & 2023 Dec 28 & Full & Detection\\
LCO-SSO-M4/MUSCAT4 & {\it $z_s$} & 2023 Dec 28 & Full & Detection\\
LCO-HAL-M3/MUSCAT3 & {\it Sloan-$g'$} & 2024 Feb 4 & Full & Detection\\
LCO-HAL-M3/MUSCAT3 & {\it Sloan-$r'$} & 2024 Feb 4 & Full & Detection\\
LCO-HAL-M3/MUSCAT3 & {\it Sloan-$i'$} & 2024 Feb 4 & Full & Detection\\
LCO-HAL-M3/MUSCAT3 & {\it $z_s$} & 2024 Feb 4 & Full & Detection\\
SSO-Europa & {\it Sloan-$r'$} & 2025 Apr 3 & Full & Detection\\
\multicolumn{5}{c}{\textbf{Spectroscopic Observations}}                                               \\ 
\textbf{Instrument} & \textbf{Wavelength Range} & \textbf{Date}     & \textbf{Number of Spectra} & \textbf{Use}\\ \midrule
Shane/Kast     & $450-900~\rm nm$      &  2025 Mar 8 & 1  & Stellar characterisation\\\midrule
IRTF/SpeX     & $800-2420~\rm nm$      &  2024 Nov 10 & 1  & Stellar characterisation\\\midrule
\end{tabular}
\label{tab:followup-6716}
\end{table*}

\begin{table*} 
\centering
\caption{Summary of ground-based follow-up observations carried out for TOI-7384.}
\begin{tabular}{@{}ccccc@{}}
\midrule \midrule
\multicolumn{5}{c}{\textbf{\textcolor{black}{TOI-7384 Follow-up Observations}}}                                              \\ \midrule \midrule
\multicolumn{5}{c}{\textbf{High Resolution Imaging}}                                             \\ 
\textbf{Observatory} & \textbf{Filter} & \textbf{Date}     & \textbf{Sensitivity Limit} & \textbf{Result}\\ \midrule
Gemini South      & $562~{\rm nm}$     &  & $\Delta m=5.6$ at $0.5\arcsec$  & No sources detected  \\ 
Gemini South      & $832~{\rm nm}$     &  & $\Delta m=5.4$ at $0.5\arcsec$  & No sources detected   \\ \midrule
\multicolumn{5}{c}{\textbf{Photometric Follow-up}}                                               \\
\textbf{Observatory} & \textbf{Filter} & \textbf{Date}     & \textbf{Coverage} & \textbf{Result} \\ \midrule
TRAPPIST-South & {\it $I+z'$} & 2021 Sept 19 & Full & Detection \\
SSO-Callisto    & {\it $I+z'$}     & 2021 Oct 14 & Full  & Detection\\
SSO-Ganymede    & {\it Sloan-$g'$}     & 2021 Oct 14 & Full  & Detection\\
TRAPPIST-South    & {\it $I+z'$}     & 2022 Apr 13 & Full  & Detection\\
SSO-Io    & {\it $I+z'$}     & 2022 Oct 23 & Full  & Detection\\
SSO-Callisto    & {\it$zYJ$} & 2022 Oct 23 & Full  & Detection\\
TRAPPIST-South    & {\it $I+z'$}     & 2022 Oct 26 & Partial  & Detection\\
SSO-Io    & {\it Sloan-$g'$} & 2022 Nov 17 & Full  & Detection\\
SSO-Europa    & {\it Sloan-$z'$} & 2022 Nov 17 & Full  & Detection\\
SSO-Callisto    & {\it$zYJ$} & 2022 Nov 17 & Full  & Detection\\
SSO-Io    & {\it $I+z'$} & 2022 Dec 12 & Full  & Detection\\
SSO-Ganymede    & {\it Sloan-$i'$} & 2023 Jan 06 & Full  & Detection\\
SSO-Io    & {\it $I+z'$} & 2023 Jan 06 & Full  & Detection\\
SSO-Europa    & {\it Sloan-$r'$} & 2023 Jan 06 & Full  & Detection\\
\multicolumn{5}{c}{\textbf{Spectroscopic Observations}}                                               \\ 
\textbf{Instrument} & \textbf{Wavelength Range} & \textbf{Date}     & \textbf{Number of Spectra} & \textbf{Use}\\ \midrule
Magellan/LDSS3     & $380-1000~\rm nm$      &  2022 Jan 06 & 1  & Stellar characterisation  \\
Magellan/FIRE     & $380-1000~\rm nm$      &  2023 Feb 02 & 1  & Stellar characterisation  \\\midrule

\end{tabular}
\label{tab:followup-7384}
\end{table*}

\subsection{Archival imaging}
In order to check for a potential blend with our target stars and background objects, we make use of various archival images available to us and compare to images taken from recent ground-based observations. These are shown in Fig.~\ref{fig:archival} for TOI-6716 (top) and TOI-7384 (bottom). This investigation is made possible by the stars' large proper motion \citep[$\mathrm{PM}_\mathrm{6716}=322 \,\mathrm{mas}\,\mathrm{yr}^{-1}$, $\mathrm{PM}_\mathrm{7384}=112 \,\mathrm{mas}\,\mathrm{yr}^{-1}$;][]{gaiaDR3cat}, and is a crucial step in assessing whether the observed transit events happen on blended sources.

For TOI-6716, we examined archival imaging spanning 69 years—from a 1956 red DSS/POSS-I plate through blue (1980) and red (1995) DSS/POSS-II/UKSTU plates and compared them with recent SPECULOOS-South observations. No background source is present at the star’s current position. 
We perform the same analysis for TOI-7384, using a DSS/POSS-II/UKSTU blue plate taken 47 years before the 2023 SPECULOOS-South images (1976), along with red (1992) and IR (1996) DSS/POSS-II/UKSTU plates, we reach the same conclusion: no background star is present at the star’s current position.

\subsection{High resolution imaging - Zorro}
A critical validation and confirmation process for transiting exoplanet observations is to use high-resolution imaging to determine if any close companions exist. The presence of a close companion star, whether truly bound or line of sight, provides `third-light' contamination of the observed transit, leading to derived properties for the exoplanet and host star that are incorrect \citep{Ciardi2015, FH2017, FH2020}. Given that nearly one-half of FGK stars are in binary or multiple star systems \citep{Matson2018} high-resolution imaging yields crucial information toward our understanding of each discovered exoplanet as well as more global information on exoplanetary formation, dynamics and evolution \citep{Howell2021}.

TOI-6716 (TIC 112115898) was observed on 2024 Jan 02  UT and TOI-7384 (TIC 192833836) was observed on 2022 October 07 UT using the Zorro speckle instrument on the Gemini South 8-m telescope \citep{Scott2021}. Zorro provides simultaneous speckle imaging in two bands (562 nm and 832 nm) with output data products including a reconstructed image with robust magnitude contrast limits on companion detections. Five sets of $1000\times0.06$ second images were obtained for TOI-6716 and 15 similar sets were obtained for TOI-7384. All these data were processed with our standard reduction pipeline \citep{Howell2011}. Fig.~\ref{fig:zorro} shows our final contrast curves and the 832 nm reconstructed speckle images for both stars. We find that TOI-6716 and TOI-7384 are both single stars with no companion brighter than 5-8 magnitudes and 5-6 magnitudes respectively below that of the target star from the Gemini Telescope 8-m telescope diffraction limit (20 mas) out to 1.2\arcsec. At the distance of TOI-6716 (d=19 pc) and TOI-7384 (d=67 pc) these angular limits correspond to spatial limits of 0.38 to 23 au and 1.34 to 80 au respectively.

\subsection{High resolution imaging - SOAR}

We also searched for stellar companions to TOI-6716 with speckle imaging on  the 4.1\,m Southern Astrophysical Research (SOAR) telescope \citep{tokovinin2018} on 2024 Jan 8 UT, observing in Cousins I-band, a similar visible bandpass as \textit{TESS}. This observation was sufficiently sensitive to obtain a 5$\sigma$ detection of a 5.0-magnitude fainter star at an angular distance of 1\arcsec from the target. More details of the observations within the SOAR \textit{TESS} survey are available in \citet{ziegler2020}. The 5~$\sigma$ detection sensitivity and speckle auto-correlation functions from the observations are shown in Fig~\ref{fig:SOAR_6716}. No nearby stars were detected within 3\arcsec of TOI-6716 in the SOAR observations.

\subsection{Photometric follow-up} \label{sec:follow-up}

\subsubsection{SPECULOOS}
The SPECULOOS Southern Observatory (SSO) is comprised of four Ritchey-Chr\'etien 1.0\,m-class telescopes installed at ESO Paranal Observatory in the Atacama desert \citep{SSOscopes}. Designed to hunt for small, habitable-zone planets orbiting ultra-cool stars \citep{gillon2018, speculoos}, three out of four telescopes are equipped with a deep-depletion Andor CCD camera with $2048 \times 2048$ 13.5-$\micron$ pixels. Each telescope has a field of view of $12\arcmin\,\times\,12\arcmin$ and a pixel scale of $0.35\arcsec$ \citep{SSOBook,ZunigaFernandez2024}. Since mid-2022, the fourth telescope of SSO has been equipped with SPIRIT (SPeculoos’ Infra-Red photometric Imager for Transits) \citep[see e.g.][]{munoz2025}, which is an InGaAs CMOS-based instrument, with a custom wide-pass filter called \textit{zYJ} \citep{spirit_pedersen2024}, designed to be optimised for observing SPECULOOS' cooler targets and to minimise the effects of precipitable water vapour (PWV) \citep{2023MNRAS.518.2661P}. This instrument has a smaller field of view of $6.7\arcmin\,\times\,5.3\arcmin$ given its detector size of $1280 \times 1024$ at 12-$\micron$ pitch. 

All SPECULOOS observations are processed in the first instance by an automatic data reduction pipeline, presented in \cite{SSOpipeline}. Successful observations of non-survey targets are then reprocessed using \textsc{prose}, a publicly available \textsc{python} framework for processing astronomical images\footnote{\url{https://github.com/lgrcia/prose}} as described in \citet{prosesoft,prosepaper}. Images are calibrated and aligned before performing aperture photometry on the 500 brightest sources detected; \textsc{prose} then performs differential photometry \citep{Broeg05} on the target star to extract the lightcurve. 

\textcolor{black}{For TOI-6716\,b, we observed one full transit with SSO/Europa on 2025 Apr 3 in the \textit{Sloan-r'} band with an exposure time of 26s. 
For TOI-7384\,b, we observed 11 full transits. 
The first observations were observed simultaneously with SSO/Callisto (in \textit{I+z'}) and SSO-Ganymede (in \textit{Sloan-g'}) on 2021 Oct 14, with exposure times of 17s and 120s respectively. 
We then observed again simultaneously on 2022 Oct 23 with SSO/Io in the \textit{I+z'} band, and SSO/Callisto in the \textit{zYJ} band, with exposure times of 17s and 20s respectively.
Our third simultaneous observation of a full TOI-7384\,b transit was on 2022 Nov 17, with SSO/Io (in \textit{Sloan-g'}, exposure time of 120s), SSO/Europa (in \textit{Sloan-z'}, exposure time of 26s) and SSO/Callisto (in \textit{zYJ}, exposure time of 20s). 
The final four transits were obtained on 2022 Dec 12 with SS0/Io (in $I+z'$, with an exposure time of 17s) and three simultaneous observations on 2023 Jan 06 with SSO/Ganymede (in $Sloan-i'$, with an exposure time of 36s), SSO/Io (in $I+z'$, with an exposure time of 17s) and SSO/Europa (in $Sloan-r'$, with an exposure time of 72s).}

\subsubsection{TRAPPIST-South}

We observed two full transits and one partial transit of TOI-7384\,b with TRAPPIST-South (TS) \citep{Jehin2011,Gillon2011}, located in ESO La Silla Observatory in Chile. This 0.6-m telescope is equipped with a FLI ProLine PL3041-BB camera and a back-illuminated CCD with a pixel size of 0.64$\arcsec$, providing a total field of view of $22\arcmin\,\times\,22\arcmin$ for an array of $2048\,\times\,2048$ pixels. TS is an f/8 Ritchey-Chr\'etien telescope on a German equatorial mount.

\textcolor{black}{The full-transit observations took place on UT 2021 Sept 19 and 2022 Apr 13, and the partial on 2022 Oct 26, with an exposure time of 120s. All transits were observed with the custom $I+z'$ filter to maximize the photometric precision.}  We reduced the images using \textsc{prose} pipeline \citep{prosepaper,prosesoft} to extract optimal light curves.

\subsubsection{LCO-CTIO-1m0}

A full transit of TOI-6716\,b was observed with the Las Cumbres Observatory Global Telescope (LCOGT; \citealt{Brown_2013}) 1.0-m network at Cerro Tololo Inter-American Observatory (CTIO). The telescope is equipped with a $4096\,\times\,4096$ SINISTRO detector, with an image scale of $0.389\arcsec$ per pixel and a FOV of $26^{\prime} \times 26^{\prime}$.
The observation was conducted on 2023 Oct 28 UT in the Sloan-$i'$ filter with an exposure time of 49s.
The image calibration was performed using the standard LCOGT {\tt BANZAI} pipeline \citep{McCully_2018SPIE10707E}. 
The closest Gaia star is TIC~777580018 at $18.24\arcsec$ with a $T_{\rm mag}$ of 16.23. We performed the aperture photometry in an uncontaminated aperture of 3.5\arcsec using {\tt AstroImageJ} \citep{Collins_2017}. 

\subsubsection{LCOGT-2m0}

Two full transits of TOI-6716\,b were observed with LCOGT-2m0 Faulkes Telescope North (FTN) at Haleakala Observatory in Hawaii (HAL) and the LCOGT-2m0 Faulkes Telescope South (FTS) at Siding Spring Observatory in Australia (SSO). The FTN telescope is equipped with the MuSCAT3 multi-band imager, while the FTS telescope is equipped with the MuSCAT4 multi-band imager \citep{Narita:2020}. The first transit was observed with MuSCAT4 on UT October 28, 2023, while the second transit was observed with MuSCAT3 on UT February 4, 2024.
The observations were conducted simultaneously in the Sloan-$g'$, -$r'$, -$i'$, and Pan-STARRS-$z_s$ filters, with exposure times of 180s, 38s, 16s, and 14s, respectively. The data processing was performed using the standard LCOGT {\tt BANZAI} pipeline \citep{McCully_2018SPIE10707E}. We performed the aperture photometry using uncontaminated apertures of 4.6--4.8\arcsec using {\tt AstroImageJ}. 

\subsection{Statistical Validation}
\label{sec:validation}
\color{black}
We make use of the statistical validation package \textsc{triceratops} \citep{triceratops_code, triceratops_paper} to validate the planetary nature of both candidates. \textcolor{black}{This follows the same procedures outlined in e.g., \citet{Dransfield2023,silverstein2024, Hesse2025, thomas2025, scottM2025}, however we outline this process again here for clarity.}

\textsc{triceratops} evaluates the likelihood that TOI-6716\,b and TOI-7384\,b are in fact true planets orbiting the target stars. \textsc{triceratops} also estimates the contributed flux from any nearby stars in order to determine whether one could be the actual source of the observed transit signal. Following this, it fits light curve models to the phase-folded photometric data in order to calculate relative probabilities for various scenarios, such as transiting planet (TP) and eclipsing binary (EB) on both the target star and nearby stars. \textsc{triceratops} then calculates a false positive probability (FPP) with a threshold for statistical validation at $\leq0.015$. 

\textcolor{black}{For TOI-6716\,b we use the two $z_s$-band light curves from MUSCAT3 and MUSCAT4, and for TOI-7384\,b we use the three $I+z'$-band light curves from SSO/Io. We find FPP$<10^{-8}$, NFPP$<10^{-7}$ and FPP$<10^{-14}$, NFPP$<10^{-17}$ for TOI-6716\,b and TOI-7384\,b respectively.}

\color{black}
\section{Global photometric analysis} \label{sec:analysis}

We use \textsc{allesfitter} \citep{AllesfitterSoft, AllesfitterPaper} to jointly model the full photometric datasets described in Section \ref{sec:follow-up} (as well as the available \textit{TESS} data) of each planet respectively. \textsc{allesfitter} is a flexible inference package for exoplanet modelling written in \textsc{Python}. It uses \textsc{ellc} \citep{ellc} to generate light curve models and \textsc{celerite} \citep{celerite} to create Gaussian Process (GP) models. \textsc{allesfitter} then uses either a nested sampling \citep[with \textsc{dynesty}][]{dynesty} or MCMC \citep[with \textsc{emcee}][]{emcee} algorithm to select the best fitting model. As it is important to be able to quantify why a specific model is favoured over another (for example, a circular vs eccentric model), we choose to use the nested sampling algorithm for this work as at each step in the sampling it calculates the Bayesian evidence from the Bayes factor \citep{BayesFactor}, allowing us to compare the evidence and determine the most statistically favoured model. This section follows process used in similar works \citep[see, e.g.,][]{Dransfield2023, scottM2025}, however we outline the steps again here for clarity. 

We adopt the transit parameters from Section \ref{sec:planet_identification} as uniform priors, and the stellar parameters described in Section \ref{sec:stellar_char} (and displayed in Table~\ref{tab:starpar}) as normal priors. The fitted parameters are $R_p/R_\star$, $(R_p+R_\star)/a$, $\cos i$, $T_0$ and $P$. Additionally, we make use of \textsc{PyLDTK} \citep{pyldtk} and Phoenix stellar atmosphere models \citep{phoenix} to calculate quadratic limb darkening coefficients. We reparameterize them following \citet{kippingldcs} and adopt them as normal priors in our fit. For observations taken in the same photometric band, the limb darkening coefficients are coupled. All prior distributions can be found in Table~\ref{tab:priors}.

We fit for two models (circular and free eccentricity) using the nested sampling algorithm. As in \citet{Triaud2011}, the eccentricity is parameterized as $\sqrt{e_b}\cos \omega_b$ and $\sqrt{e_b}\sin \omega_b$. We calculate the difference in the log marginal likelihoods (Bayes factor), $\Delta\log Z$, in order to determine which model is favoured. Here, we adopt the circular fit as the null hypothesis, $H_0$, and the eccentric fit as the alternate hypothesis, $H_1$. $H_1$ is preferred over $H_0$ if $\Delta\log Z > 3$, and strongly preferred if $\Delta\log Z > 5$ \citep{Trotta2008}.
We find that for TOI-6716\,b, there is no strong preference for either model, with $\Delta\log Z=-0.4$, and similarly for TOI-7384\,b, with $\Delta\log Z=-1.5$. However, since eccentricity cannot be well constrained from photometry alone, we do not assume that the orbits of these planets are perfectly circular, and thus adopt the eccentric model fit parameters where we place an upper limit (95\% confidence) on the eccentricities of both planets to be $e\leq0.88$ and $e\leq0.40$ for TOI-6716\,b and TOI-7384\,b respectively. We note that our derived results have no significant change between circular and eccentric fits.

\begin{figure}
    \centering
    \includegraphics[width=1\columnwidth]{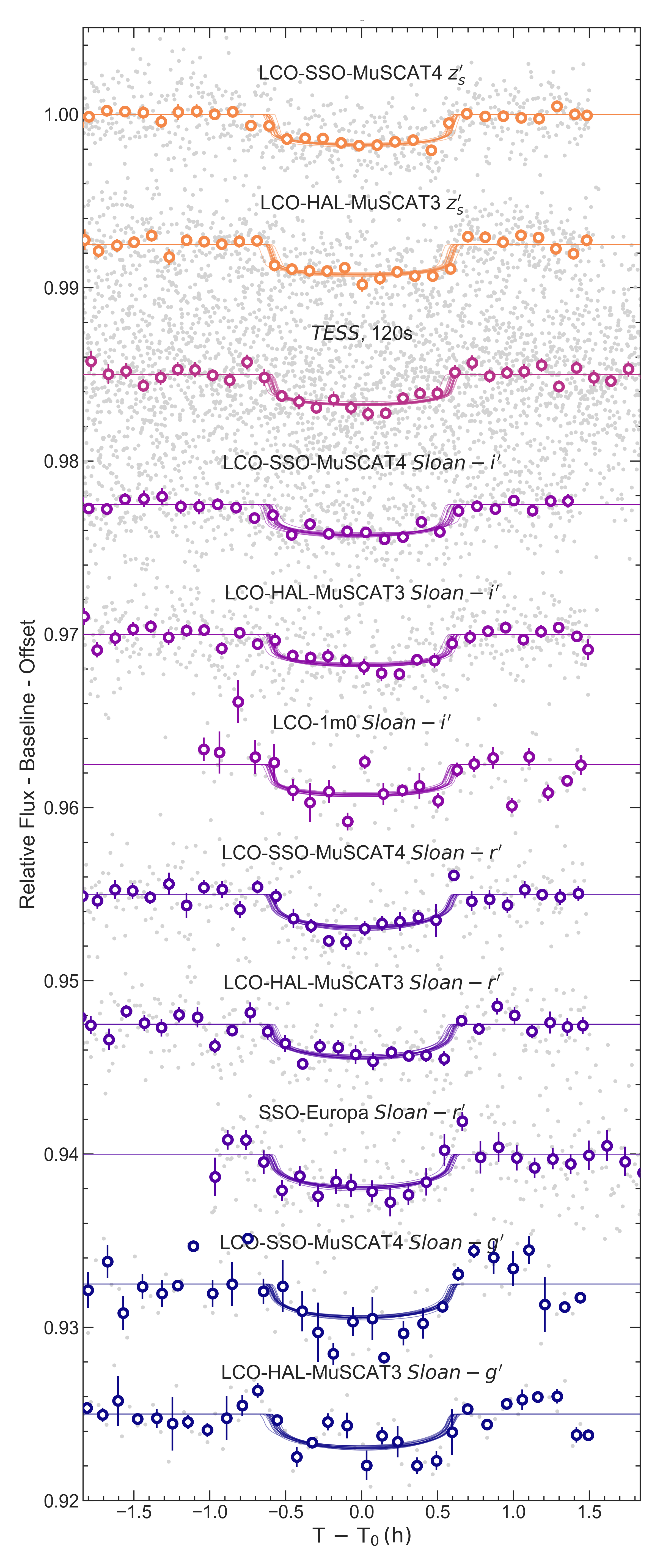}
    \caption{Phase-folded transits of TOI-6716\,b from \textit{TESS} 2-minute cadence and follow-up, ground-based observations. Raw flux points are shown in grey and binned (8 min) in white circles. Transit models are corrected by subtracting the baseline, and all have a relative offset applied for plotting purposes. Model lines comprise of 20 random draws from the posterior transit model.}
    \label{fig:transits_6716}
\end{figure}

\begin{figure}
    \centering
    \includegraphics[width=1\columnwidth]{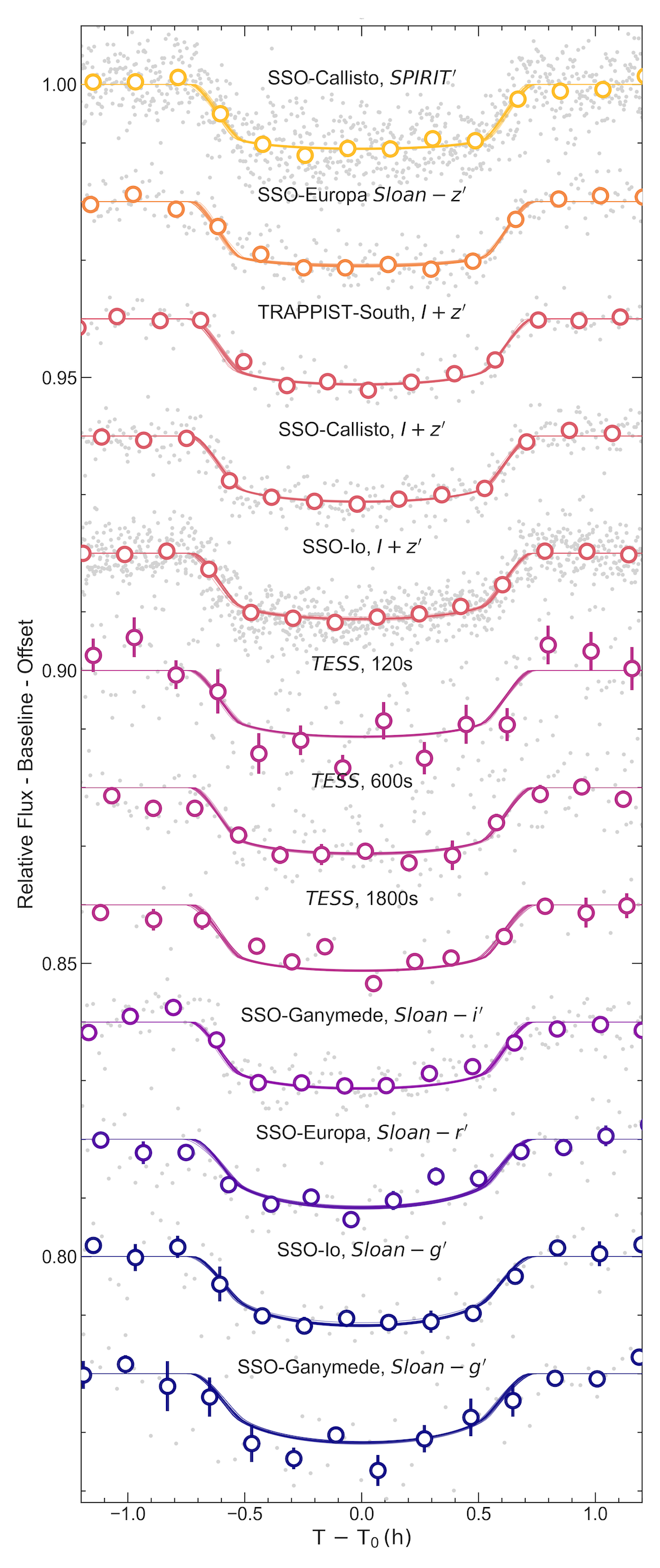}
    \caption{Phase-folded transits of TOI-7384\,b from \textit{TESS} 2-,10-, and 30-minute cadence and follow-up, ground-based observations. Raw flux points are shown in grey and binned (10 min) in white circles. Transit models are corrected by subtracting the baseline, and all have a relative offset applied for plotting purposes. Model lines comprise of 20 random draws from the posterior transit model.}
    \label{fig:transits_7384}
\end{figure}

In Fig.~\ref{fig:transits_6716} and Fig.~\ref{fig:transits_7384} we present the transits obtained for TOI-6716\,b and TOI-7348\,b respectively. Where multiple transits were observed with the same instrument and filter, the transits are presented phase-folded. All fitted and derived parameters are presented in Tables~\ref{tab:fitted_results} and ~\ref{tab:derived_results}. 

\begin{figure*} 
  \centering
  \subfloat[TOI-6716\,b averaged depths with dilution from chromatic fit.]{\includegraphics[width=\columnwidth]{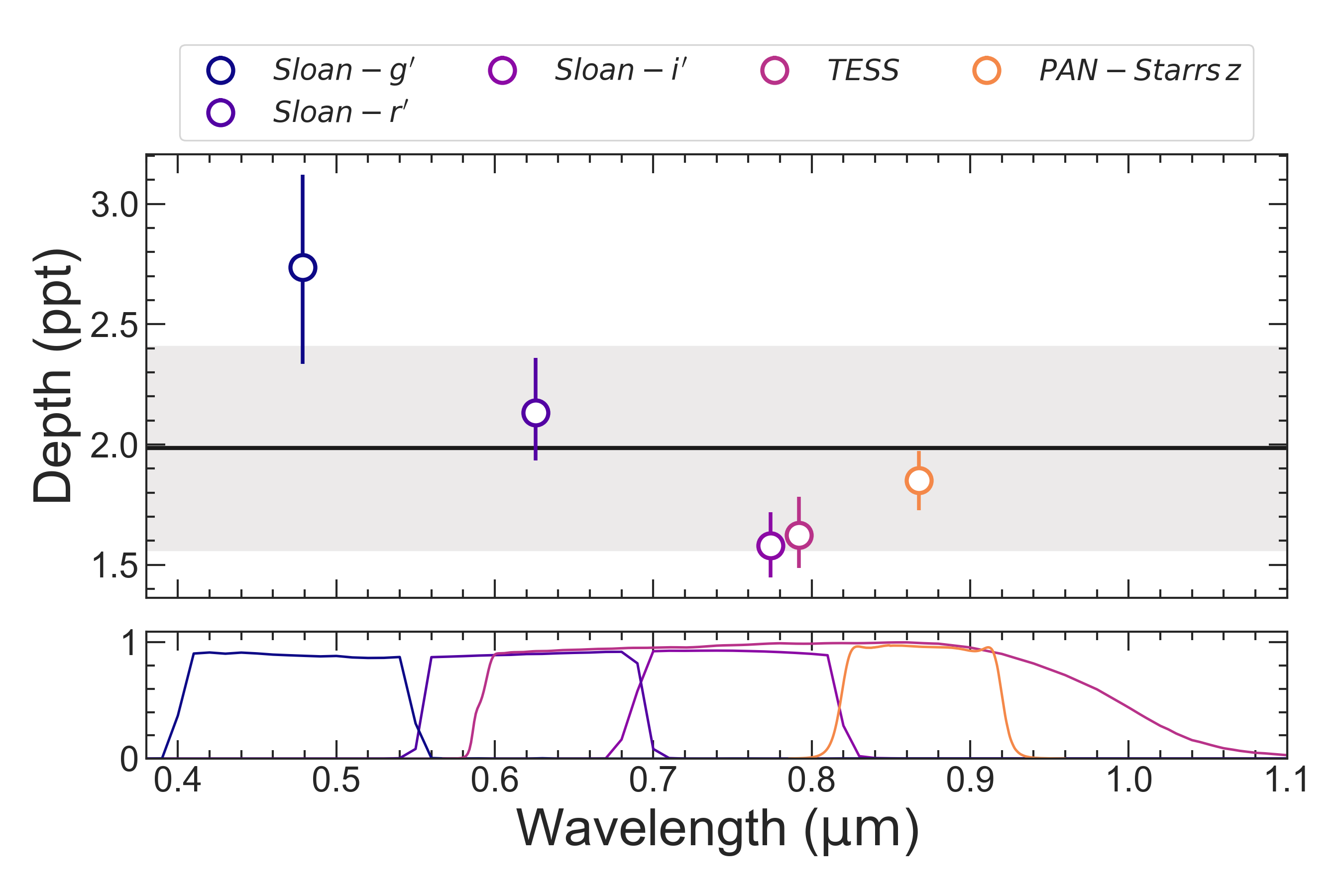}}\label{fig:6716_chrom}\quad 
  \subfloat[TOI-7384\,b averaged depths with dilution from chromatic fit.]{\includegraphics[width=\columnwidth]{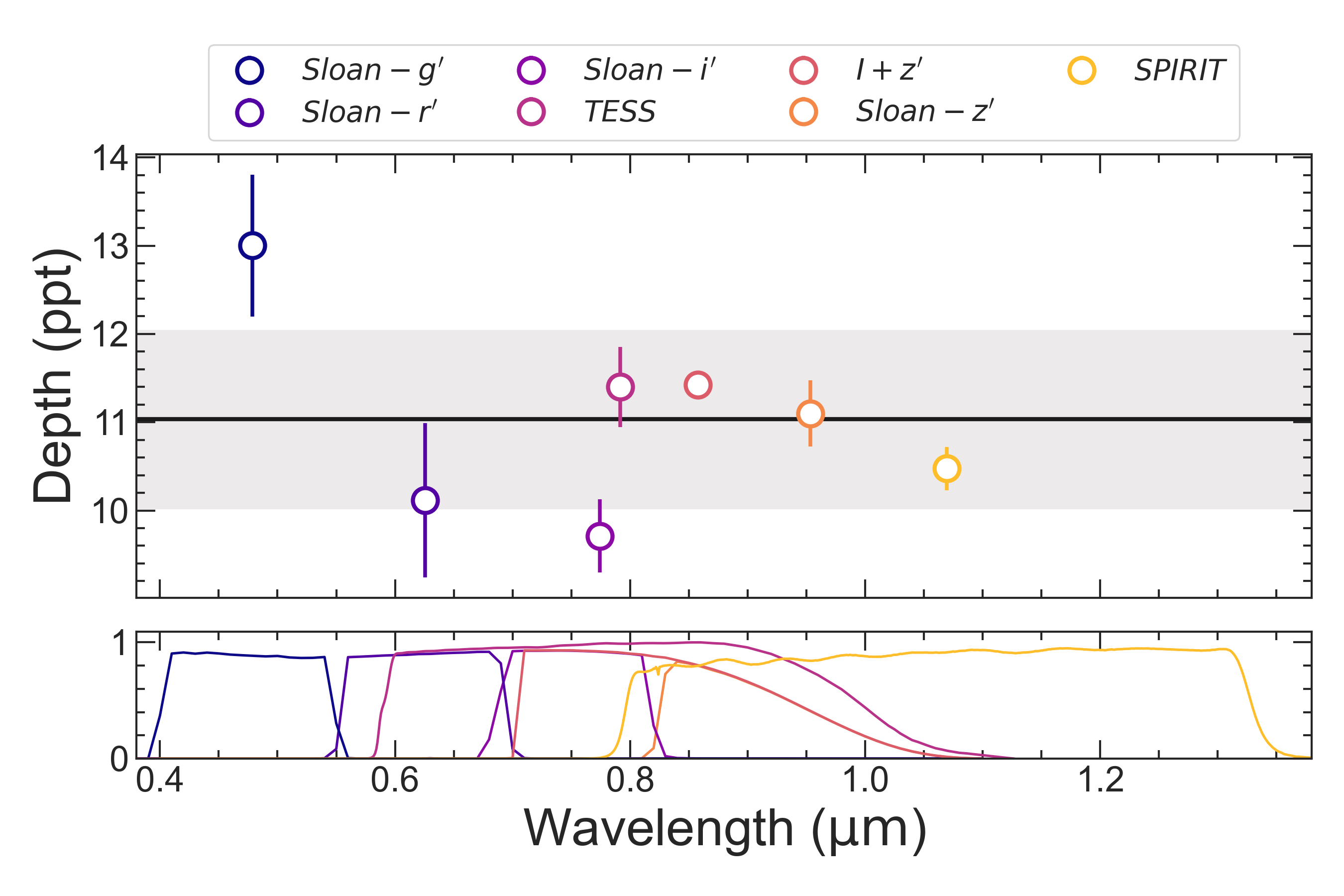}} \label{fig:7384_chrom}
  \caption{Chromaticity check of TOI-6716\,b (left) and TOI-7384\,b (right). Depths are obtained from the chromatic fit and are corrected for limb-darkening. Grey band shows the mean depth and coloured points show the average depths in each band. For TOI-6716\,b, all bands agree to within 1.3$\sigma$. For TOI-7384\,b, all bands agree to within 1.6$\sigma$.}
  \label{fig:chromaticity_check}
\end{figure*}

As an additional validation check, we perform the fit again using the priors in Table~\ref{tab:priors}, however we allow free dilution ($\mathcal{U}[-1, 1]$) in order to check for chromaticity. Dilution is coupled between observations taken in the same band. \textsc{allesfitter} provides the fitted diluted minimum in-transit flux from each instrument, which we then correct with the calculated impact parameter and respective limb darkening coefficients to obtain $R_\mathrm{p}/R_\star$ in each band. If TOI-6716\,b and TOI-7384\,b are, in fact, planets (rather than, e.g., eclipsing binaries), we would expect consistent depths across the photometric bands they were observed in, i.e., achromaticity. In Fig.~\ref{fig:chromaticity_check}, we present these averaged depths for both targets transits, and show that all depths are consistent to 1$\sigma$.  

\textcolor{black}{Finally, we also compare the stellar host densities calculated by \textsc{allesfitter} from the transit parameters \citep[following][]{Seager2003} to that of the stellar density priors calculated from the SED parameters for each star respectively. For both TOI-6716 and TOI-7384, we find these values to be within $1\sigma$ of the stellar density priors ($25.44\pm5.11$ and $13.78\pm2.43$ respectively), suggesting the planetary transit signal is produced over the target star.}

\begin{table} 
\centering
\caption{Fitted parameters from achromatic eccentric fits for TOI-6716\,b. and TOI-7384\,b.}
\begin{tabular}{@{}ccccc@{}} 
\midrule \midrule
Parameter & 6716\,b & 7384\,b  \\ 
\hline 
\multicolumn{3}{c}{\textit{Fitted parameters}} \\ 
\hline 
$R_b / R_\star$ & $0.03898_{-0.00075}^{+0.00081}$ & $0.10238\pm0.00082$ \\
$(R_\star + R_b) / a_b$ & $0.0348_{-0.0023}^{+0.0027}$ & $0.0371_{-0.0018}^{+0.0023}$ \\
$\cos{i_b}$ & $0.0096_{-0.0062}^{+0.0072}$ & $0.0211_{-0.0023}^{+0.0032}$ \\
$T_{0;b}$ ($\mathrm{BJD}$) & $2459632.34484_{-0.00086}^{+0.00062}$ & $2459565.10114_{-0.00017}^{+0.00018}$ \\
$P_b$ ($\mathrm{d}$) & $4.7185898_{-0.0000041}^{+0.0000054}$ & $6.2340258_{-0.0000036}^{+0.0000034}$ \\
$\sqrt{e_b} \cos{\omega_b}$ & $-0.04_{-0.62}^{+0.48}$ & $0.03_{-0.37}^{+0.39}$ \\
$\sqrt{e_b} \sin{\omega_b}$ & $-0.16\pm0.23$ & $-0.00_{-0.16}^{+0.18}$ \\
$q_{1; \mathrm{TESS}}$ & $0.302_{-0.043}^{+0.046}$ & $0.2853\pm0.0045$ \\
$q_{2; \mathrm{TESS}}$ & $0.309_{-0.042}^{+0.040}$ & $0.3044\pm0.0049$ \\
$q_{1; \mathrm{{i'}}}$ & $0.395_{-0.043}^{+0.041}$ & $0.3431\pm0.0047$ \\
$q_{2; \mathrm{{i'}}}$ & $0.294_{-0.043}^{+0.039}$ & $0.3083\pm0.0046$ \\
$q_{1; \mathrm{{g'}}}$ & $0.760_{-0.043}^{+0.048}$ & $0.6875\pm0.0044$ \\
$q_{2; \mathrm{{g'}}}$ & $0.323\pm0.046$ & $0.3585\pm0.0047$ \\
$q_{1; \mathrm{{r'}}}$ & $0.685\pm0.042$ & $0.5983\pm0.0044$ \\
$q_{2; \mathrm{{r'}}}$ & $0.361\pm0.040$ & $0.3772\pm0.0045$ \\
$q_{1; \mathrm{{z_s}}}$ & $0.270\pm0.042$ & - \\
$q_{2; \mathrm{{z_s}}}$ & $0.261\pm0.039$ & - \\
$q_{1; \mathrm{{I+z'}}}$ & - & $0.2644\pm0.0046$  \\ 
$q_{2; \mathrm{{I+z'}}}$ & - & $0.2884\pm0.0045$ & \\ 
$q_{1; \mathrm{{z'}}}$ & - & $0.1288\pm0.0045$ \\ 
$q_{2; \mathrm{{z'}}}$ & - &  $0.2881\pm0.0046$ \\ 
$q_{1; \mathrm{spirit}}$ & - & $0.1338\pm0.0044$ \\ 
$q_{2; \mathrm{spirit}}$ & - & $0.2874\pm0.0044$ \\ 
$\ln{\sigma_\mathrm{TESS_{120s}}}$ & $-5.5237\pm0.0068$ & $-4.300\pm0.018$ \\
$\ln{\sigma_\mathrm{TESS_{600}}}$ & - & $-5.408\pm0.026$ \\ 
$\ln{\sigma_\mathrm{TESS_{1800}}}$ & - & $-5.850\pm0.050$ \\
$\ln{\sigma_\mathrm{LCO1m0_{i'}}}$ & $-6.205_{-0.058}^{+0.068}$ & - \\
$\ln{\sigma_\mathrm{LCO-SSO-M4_{i'}}}$ & $-6.492\pm0.027$ & - \\
$\ln{\sigma_\mathrm{LCO-Hal-M3_{i'}}}$ & $-6.377_{-0.030}^{+0.032}$ & - \\
$\ln{\sigma_\mathrm{ganymede_{i'}}}$ & - & $-5.735_{-0.026}^{+0.025}$ & \\ 
$\ln{\sigma_\mathrm{LCO-SSO-M4_{g'}}}$ & $-6.281\pm0.083$ & - \\
$\ln{\sigma_\mathrm{LCO-Hal-M3_{g'}}}$ & $-6.669_{-0.072}^{+0.087}$ & - \\
$\ln{\sigma_\mathrm{ganymede_{g'}}}$ & - & $-5.094_{-0.049}^{+0.053}$ \\ 
$\ln{\sigma_\mathrm{io_{g'}}}$ & - & $-5.562_{-0.049}^{+0.053}$ \\ 
$\ln{\sigma_\mathrm{europa_{r'}}}$ & $-5.948\pm0.039$ & $-5.337\pm0.035$ \\ 
$\ln{\sigma_\mathrm{LCO-SSO-M4_{r'}}}$ & $-6.433_{-0.037}^{+0.042}$ & - \\
$\ln{\sigma_\mathrm{LCO-Hal-M3_{r'}}}$ & $-6.488_{-0.037}^{+0.039}$ & - \\
$\ln{\sigma_\mathrm{LCO-SSO-M4_{z_s}}}$ & $-6.544_{-0.021}^{+0.026}$ & - \\
$\ln{\sigma_\mathrm{LCO-Hal-M3_{z_s}}}$ & $-6.434\pm0.024$ & - \\
$\ln{\sigma_\mathrm{Io_{I+z'}}}$ & - & $-5.712\pm0.013$\\ 
$\ln{\sigma_\mathrm{TRAPPIST\,SOUTH_{I+z'}}}$ & - & $-5.869_{-0.048}^{+0.051}$\\ 
$\ln{\sigma_\mathrm{callisto_{I+z'}}}$ & - & $-5.855\pm0.024$ \\ 
$\ln{\sigma_\mathrm{europa_{z'}}}$ & - & $-5.982\pm0.029$ \\ 
$\ln{\sigma_\mathrm{spirit}}$ & - & $-5.370\pm0.014$ \\ 

\end{tabular} \label{tab:fitted_results}
\end{table}

\begin{table*}
\centering
\caption{Derived parameters from achromatic eccentric fit for TOI-6716\,b and TOI-7384\,b. Since eccentricity is unconstrained, we quote here the 95\% confidence upper limit. We thus do not report $\omega$ as it was also unconstrained.}
\begin{tabular}{@{}ccccc@{}}
\midrule \midrule
Parameter & 6716\,b & 7384\,b  \\ 
\hline
\multicolumn{3}{c}{\textit{Derived parameters}} \\ 
\hline 
Host radius over semi-major axis b; $R_\star/a_\mathrm{b}$ & $0.0335_{-0.0022}^{+0.0027}$ & $0.0337_{-0.0017}^{+0.0021}$ \\ 
Semi-major axis b over host radius; $a_\mathrm{b}/R_\star$ & $29.9\pm2.2$ & $29.7_{-1.8}^{+1.6}$ \\ 
Companion radius b over semi-major axis b; $R_\mathrm{b}/a_\mathrm{b}$ & $0.001304_{-0.000087}^{+0.00011}$ & $0.00345_{-0.00018}^{+0.00022}$ \\ 
Companion radius b; $R_\mathrm{b}$ ($\mathrm{R_{\oplus}}$) & $0.982\pm0.067$ & $3.56\pm0.21$ \\ 
Companion radius b; $R_\mathrm{b}$ ($\mathrm{R_{jup}}$) & $0.0876\pm0.0060$ & $0.318\pm0.018$ \\ 
Semi-major axis b; $a_\mathrm{b}$ ($\mathrm{R_{\odot}}$) & $6.87\pm0.67$ & $9.45\pm0.78$ \\ 
Semi-major axis b; $a_\mathrm{b}$ (AU) & $0.0319\pm0.0031$ & $0.0439\pm0.0036$ \\ 
Inclination b; $i_\mathrm{b}$ (deg) & $89.45_{-0.42}^{+0.36}$ & $88.79_{-0.18}^{+0.13}$ \\ 
Eccentricity b; $e_\mathrm{b}$ & $\leq0.88$ & $\leq0.40$ \\ 
Impact parameter b; $b_\mathrm{tra;b}$ & $0.27\pm0.17$ & $0.611_{-0.037}^{+0.031}$ \\ 
Total transit duration b; $T_\mathrm{tot;b}$ (h) & $1.237\pm0.015$ & $1.452\pm0.015$ \\ 
Full-transit duration b; $T_\mathrm{full;b}$ (h) & $1.135_{-0.016}^{+0.014}$ & $1.041\pm0.018$ \\ 
Host density from orbit b; $\rho_\mathrm{\star;b}$ (cgs) & $22.6_{-4.6}^{+5.1}$ & $12.8\pm2.2$ \\ 
Equilibrium temperature b; $T_\mathrm{eq;b}$ (K) & $369_{-16}^{+17}$ & $378_{-13}^{+15}$ \\ 
\textcolor{black}{Instellation Flux b; $S_{\mathrm{b}}$ ($\mathrm{S}_\oplus$)} & \textcolor{black}{\instellationA} & \textcolor{black}{\instellationB} \\
Limb darkening; $u_\mathrm{1; TESS}$ & $0.338\pm0.059$ & $0.3251\pm0.0058$ \\ 
Limb darkening; $u_\mathrm{2; TESS}$ & $0.210\pm0.044$ & $0.2089\pm0.0055$ \\ 
Limb darkening; $u_\mathrm{1; {i'}}$ & $0.368_{-0.058}^{+0.052}$ & $0.3611\pm0.0064$ \\ 
Limb darkening; $u_\mathrm{2; {i'}}$ & $0.258_{-0.051}^{+0.057}$ & $0.2246\pm0.0054$ \\ 
Limb darkening; $u_\mathrm{1; {g'}}$ & $0.564\pm0.081$ & $0.5944_{-0.0076}^{+0.0082}$ \\ 
Limb darkening; $u_\mathrm{2; {g'}}$ & $0.310\pm0.083$ & $0.2346\pm0.0077$ \\ 
Limb darkening; $u_\mathrm{1; {r'}}$ & $0.597\pm0.073$ & $0.5834\pm0.0071$ \\ 
Limb darkening; $u_\mathrm{2; {r'}}$ & $0.230\pm0.067$ & $0.1899\pm0.0072$ \\ 
Limb darkening; $u_\mathrm{1; {z_s}}$ & $0.269_{-0.048}^{+0.053}$ & - \\ 
Limb darkening; $u_\mathrm{2; {z_s}}$ & $0.246_{-0.042}^{+0.045}$ & - \\ 
Limb darkening; $u_\mathrm{1; {I+z'}}$ & - & $0.2966\pm0.0055$  \\ 
Limb darkening; $u_\mathrm{2; {I+z'}}$ & - & $0.2176\pm0.0050$ \\ 
Limb darkening; $u_\mathrm{1; {z'}}$ & - & $0.2069_{-0.0050}^{+0.0046}$ \\ 
Limb darkening; $u_\mathrm{2; {z'}}$ & - & $0.1521\pm0.0044$ \\ 
Limb darkening; $u_\mathrm{1; spirit}$ & - & $0.2103\pm0.0049$ \\ 
Limb darkening; $u_\mathrm{2; spirit}$ & - & $0.1554\pm0.0042$ \\ 
Combined host density from all orbits; $\rho_\mathrm{\star; combined}$ (cgs) & $22.6_{-4.6}^{+5.1}$ & $12.8\pm2.2$ \\ 
\end{tabular} \label{tab:derived_results}
\end{table*}

\section{Discussion and conclusion} \label{sec:discussion}

\begin{figure*}
    \centering
    \includegraphics[scale=0.7]{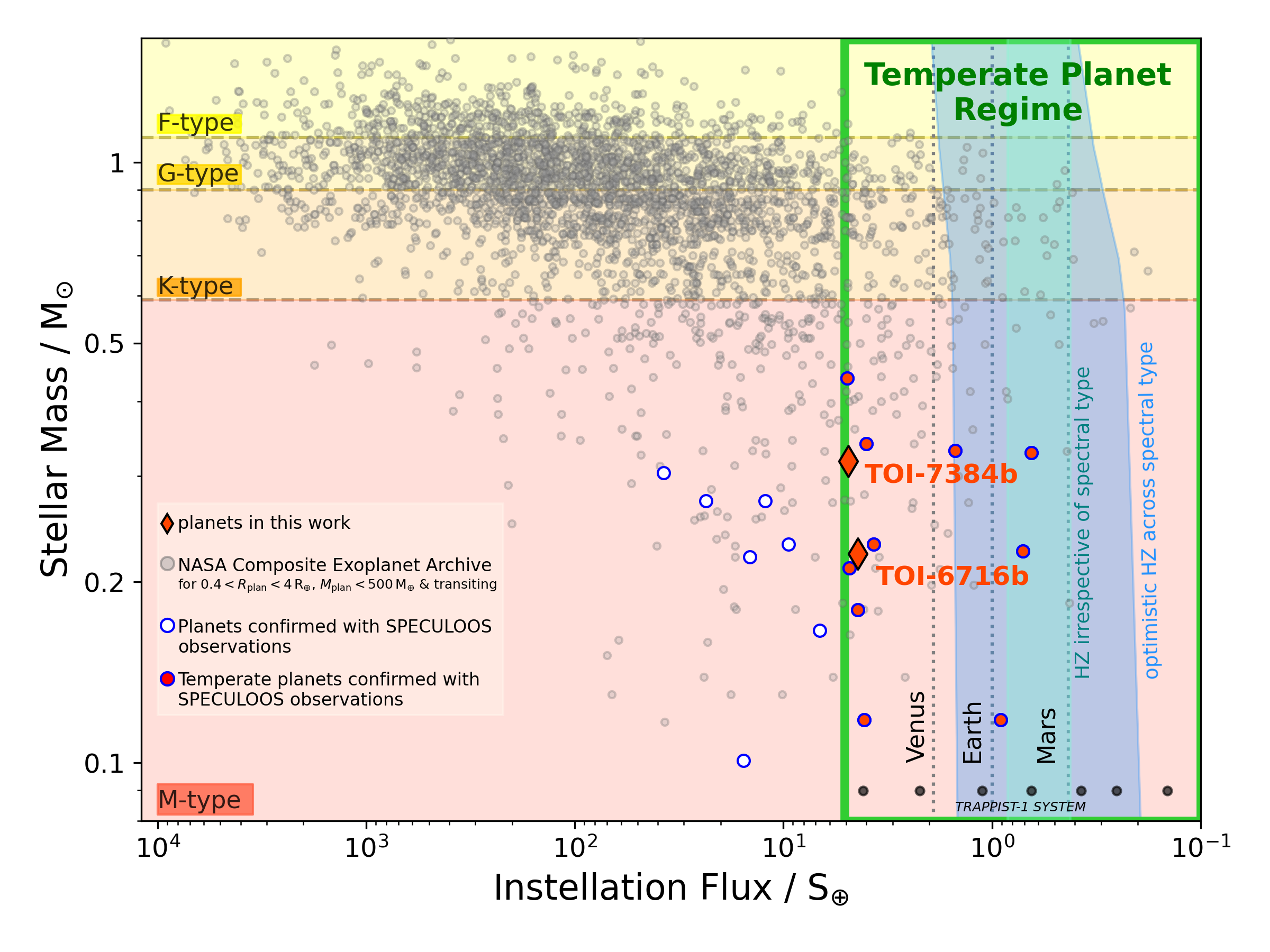}
    \caption{Stellar mass vs instellation flux for transiting planets with $0.4<R_p<4\,\mathrm{R}_\oplus$ and $M_p<500\,\mathrm{M}_\oplus$ obtained from the Composite NASA exoplanet archive (grey points). Instellation flux recalculated for all planets for consistency. Approximate spectral types are indicated by the shaded background and highlighted labels. Planets for which SPECULOOS observations contributed to the validation/confirmation are highlighted with blue circles. TOI-6716\,b and TOI-7384\,b are depicted by red diamonds, and sit at the inner (hotter) edge of the temperate regime (shown by the green box).}
    \label{fig:mass-instellation}
\end{figure*}

TOI-6716 and TOI-7384 join the select club of mid-type, fully convective M-dwarfs known to host temperate planets. TOI-6716\,b is an Earth-sized,\radiusA exoplanet orbiting with a period of\periodA, thus receiving an instellation flux of \instellationA. This places it near the inner edge of the temperate zone. TOI-7384\,b is a Neptune-sized exoplanet, with \radiusB, that orbits with a period of \periodB and receives a similar instellation flux of \instellationB.
While neither of these planets fall even within the \textit{optimistic} HZs of their stars \citep[orbital periods between 9.9-43.1\,days and 14.4-62.5\,days for TOI-6716 and TOI-7384 respectively;][]{2013ApJ...765..131K}, these planets populate an otherwise sparse region of temperate planet parameter space, offering opportunities for future studies as definitions of exoplanet habitability broaden. 

In this section we describe TOI-6176\,b and TOI-7384\,b within the current population of exoplanets, and then describe briefly the future prospects for follow-up studies.

\subsection{TOI-6716\,b and TOI-7384\,b within the current population}

\subsubsection{Stellar mass and instellation flux}
Fig.~\ref{fig:mass-instellation} shows TOI-6716\,b and TOI-7384\,b within the current population of exoplanets obtained from the NASA Exoplanet Archive's Composite Database\footnote{Retrieved on 2025 Jul 15, \url{https://exoplanetarchive.ipac.caltech.edu/cgi-bin/TblView/nph-tblView?app=ExoTbls&config=PSCompPars}}, for transiting planets with \textcolor{black}{$0.4<R_\mathrm{p}<4\,\mathrm{R}_\oplus$} and $M_{\rm p}<500\,\mathrm{M}_\oplus$. For consistency, we recalculate all planetary instellation fluxes.

The green box highlights planets which fall within the temperate regime, where we see that TOI-6716\,b and TOI-7384\,b are at the inner (hotter) edge. We also highlight those planets which had SPECULOOS observations that aided in their validation/confirmation, for which the planets in this work also add to this sample.

\subsection{Future prospects}
\color{black}
\subsubsection{More precise planetary radii?}
The goal of the TEMPOS programme is to achieve very precise temperate exoplanet radii, ideally $\leq3\%$ (see Section~\ref{sec:tempos}). 
Currently, TOI-6716\,b and TOI-7384\,b have radii precisions of $6.8\%$ and $5.9\%$ respectively, both of which are dominated by their stellar radius precisions (which set planet radius error floors at $6.5\%$ and $5.7\%$). 
Thus in order to significantly improve the planet radius error the stellar radius error must decrease. \textcolor{black}{
Surveys such as the EBLM (Eclipsing Binaries-Low Mass) project \citep{triaud2013, vonboetticher2019,maxted2023,swayne2024} are aiming to create an empirical mass-radius-metallicity-luminosity relation using single-lined eclipsing binaries and recently some double-lined systems \citep{sebastian2024,sebastian2025,baycroft2025,triaud2025}. An new analysis of the project's result recently achieved a radius accuracy of 1.4\% (Davis et al., in prep) across dozens of systems. When these binary star results will have been used to derive new mass/radius relations, we expect to achieve accurate and precise masses and radii for low-mass field M dwarfs, which when included into the TEMPOS program will aid in reaching our planetary radius precision goal ($\leq3\%$).
}
For TOI-6716\,b, we estimate that in order to reach a planet radius precision of $\leq3\%$ with a reasonable number of additional observations (e.g., 10), we require that the stellar radius precision increase to 2.32\% (assuming a precision of $450\,$ppm on the transit depth for a single transit with SPECULOOS). 
Similarly for TOI-7384\,b, we require that the stellar radius precision increase to 2.9\%, assuming a precision of $500\,$ppm on the transit depth for a single transit with SPECULOOS.

\color{black}
\subsubsection{Planetary mass and atmospheric prospects}
In order to fully characterise a planet it is crucial to obtain its density, something which requires both photometric and spectroscopic observations. For small, Earth-sized planets, however, obtaining precise radial velocities often proves challenging, especially around late-type M dwarfs given the stellar contamination \citep[see e.g.,][]{turner2025}. Using the mass--radius relations from \citet{2017ApJ...834...17C}, we predict the masses of TOI-6716\,b and TOI-7384\,b to be $0.9\pm0.23\,\mathrm{M}_\oplus$ and $12.4\pm1.2\,\mathrm{M}_\oplus$, leading to predicted radial velocity semi-amplitudes\footnote{calculated from $K=\big(\frac{2\pi G}{P}\big)^{1/3} \frac{M_\mathrm{P}\sin(i)}{M_\star^{2/3}}$, where $M_\mathrm{P},M_\mathrm{\star}$ are the planet and star masses, $P$ is the orbital period, $i$ is the inclination, and $G$ is the gravitational constant.} of $0.9\pm0.25\,\mathrm{m}\,\mathrm{s}^{-1}$ and $9.0\pm1.2\,\mathrm{m}\,\mathrm{s}^{-1}$ respectively. 

From the MAROON-X exposure time calculator \textcolor{black}{\citep{seifahrt2018}}, we could achieve an RV precision of $1.8\,\rm m\,s^{-1}$ per measurement with an exposure time of 1800\,s at an SNR$\sim120$ \textcolor{black}{for TOI-6716}. Therefore, for a \textcolor{black}{$5\sigma$} measurement on the mass \textcolor{black}{(i.e. $\sim20\%$ precision)} \textcolor{black}{of TOI-6716\,b, $\sim200$} spectra would need to be collected, \textcolor{black}{where we assume that only approximately $2/3$ measurements truly contributed to the mass measurement}.
For TOI-7384, we find an RV precision of $11.6\,\rm m\,s^{-1}$, and with an expected SNR$\sim 23$, we would require \textcolor{black}{$\sim100$ spectra for a $5\sigma$} measurement on the mass \textcolor{black}{of TOI-7384\,b}. We assume here that both stars rotate relatively slowly, i.e. $v\sin i \lesssim2\,\rm km\, s^{-1}$, an assumption that can be made for old, relatively inactive M dwarfs \citep{moutou2017, reiners2022}

\textcolor{black}{We also repeat this calculation for observations with NIRPS \citep{bouchy2025}, where we calculate the expected precision on a singular measurement for an M4 star using,} 

\begin{equation}
    \sigma_{\mathrm{RV}} = 0.42\times10^{(0.94\,\mathrm{H}_{\mathrm{mag}}-5)/5},
\end{equation}
\color{black}
\citep{artigau2024,bouchy2025}. We find precisions of $2.99\,\rm m\,s^{-1}$ and $6.65\,\rm m\,s^{-1}$ for TOI-6716 and TOI-7384 respectively for an exposure time of 1800s. Thus for a $5\,\sigma$ mass measurement, we would require $\sim$600 measurements for TOI-6716\,b, and $\sim$30 measurements for TOI-7384\,b. 

ESPRESSO \citep{pepe2021} is another high-resolution spectrograph that could obtain RV measurements for our planetary mass measurements. However, from the ESPRESSO ETC, we calculate that for a $5\,\sigma$ mass measurement, we would require $\sim$770 and $\sim$70 measurements of TOI-6716 and TOI-7384, assuming RV precisions of $3.3\,\rm m\,s^{-1}$ and $10\,\rm m\,s^{-1}$ respectively, therefore making this facility less favourable compared to NIRPS or MAROON-X for these science goals.

While for TOI-7384\,b a fairly reasonable number of measurements are required to make a significant mass measurement (but still difficult to obtain on competitive facilities), this is not the case for small, likely Earth-mass planet TOI-6716\,b, where a much larger amount of telescope time would be required.

\color{black}
\textcolor{black}{
We emphasise that these RV estimates are based on the assumption that the predicted mass is correct, however this is not always the case and can sometimes be largely over- or under-predicted, which naturally will affect the number of RV measurements required (specifically for the under-predicted case). A recent example of this is TOI-6478\,b \citep{scottM2025}, who calculated that based on the predicted mass of the planet, 10 RVs from MAROON-X would be sufficient to obtain a $3-4\sigma$ constraint. However, the planet is seemingly under-dense, and as such those 10 RVs could not constrain the mass at all, and only provided an upper mass limit.}

Naturally, the next step in understanding these planets could be observations of their atmospheres. \citet{kempton2018} define a transmission spectroscopy metric (TSM) which is proportional to the SNR expected from the strength of possible transit features. Large TSMs are indicative of strong features, thus facilitating more detailed and accurate atmospheric characterisation. For TOI-6716\,b we predict TSM $\sim13$ (similar to the outer TRAPPIST-1 planets), and for TOI-7384\,b TSM $\sim69$. An ongoing question, however, is whether small, rocky planets around M dwarfs can retain their atmospheres; studies into the M dwarf cosmic shoreline aim to inform this debate, allowing for more targeted follow-up with, e.g., JWST \citep[see e.g.,][]{xue2025}. \citet{pass2025} investigate the cosmic shoreline for late-type M dwarfs, and calculate the cumulative historic x-ray and ultraviolet (XUV) irradiation received by a planet, $I_{XUV}$, as a function of stellar mass. Figure 2 in their work shows $I_{XUV,\oplus}$ vs escape velocity, $v_{\mathrm{esc}}$, with regions of atmospheric loss and retention and planets with $R_{\rm p}<1.8\,\mathrm{R}_\oplus$.

We calculate $v_{\mathrm{esc}} = \sqrt{\frac{M_{\rm P}}{\mathrm{M_\oplus}} \frac{\mathrm{R_\oplus}}{R_{\rm P}}} v_{\mathrm{esc, \oplus}}$ and find $v_{\mathrm{esc}}=0.96\,v_{\mathrm{esc, \oplus}}$ ($=10.7\,\mathrm{km}\,\mathrm{s}^{-1}$) for TOI-6716\,b. We then estimate $I_{\rm XUV}=422\,I_{\rm XUV,\oplus}$ using values in Table 1\footnote{We take values for TOI-6716 from the $M_\star=0.25\,\mathrm{M_\odot}$ row in Table 1 from \citet{pass2025}} from \citet{pass2025}, placing it within the region of atmospheric loss. \citet{pass2025} also define an ``atmospheric retention metric'' (ARM), which quantifies the position of the planet relative to the cosmic shoreline, for which we find ARM$\sim-1.66$. Therefore, we expect that TOI-6716\,b is unlikely to have retained its atmosphere; however, this is likely not the case for the more massive TOI-7384\,b, making it an ideal target for future JWST observations. 

\subsection{Summary}

We report the discovery and validation of two temperate planets --- an Earth-sized planet, TOI-6716\,b and a Neptune-sized planet, TOI-7384\,b --- transiting fully convective M dwarfs. These planets add to the small but growing sample of planets in the temperate region that we have defined in this work, $0.1< S <5\,\mathrm{S}_\oplus$. 

The planets were initially identified with \textit{TESS} and validated through an extensive campaign of high-resolution imaging, ground-based transit observations with SPECULOOS, TRAPPIST-South, and LCOGT, as well as reconnaissance spectroscopy of the host stars.
Statistical analysis with \textsc{triceratops} yields false positive probabilities of ${<}10^{-8}$ and ${<}10^{-14}$ for TOI-6716\,b and TOI-7384\,b, respectively.
Injection--recovery tests show that additional short-period planets larger than $\sim1\,R_\oplus$ can be excluded for TOI-6716, while Earth-sized companions remain undetectable for the current data on TOI-7384.

Given its small radius and high irradiation, TOI-6716\,b may be airless, \textcolor{black}{however with a predicted TSM similar to that of the outer TRAPPIST-1 planets, should it have an atmosphere it could be an good target for rocky-world atmospheric observations with JWST. In terms of TOI-7384\,b, its size and predicted mass make it an attractive target for atmospheric characterization with JWST and future facilities. }
\textcolor{black}{Together these discoveries show the power of combining \textit{TESS} with coordinated ground-based efforts to build a catalogue of temperate planets around fully convective M dwarfs for atmospheric studies in the coming decade.}

\section*{Acknowledgements}
MGS acknowledges support from the UK Science and Technology Facilities Council (STFC) \textcolor{black}{and from a local studentship delivered by the College of Engineering and Physical Sciences of the University of Birmingham.}

The ULiege's contribution to SPECULOOS has received funding from the European Research Council under the European Union's Seventh Framework Programme (FP/2007-2013) (grant Agreement n$^\circ$ 336480/SPECULOOS), from the Balzan Prize and Francqui Foundations, from the Belgian Scientific Research Foundation (F.R.S.-FNRS; grant n$^\circ$ T.0109.20), from the University of Liege, and from the ARC grant for Concerted Research Actions financed by the Wallonia-Brussels Federation.  This work is supported by a grant from the Simons Foundation (PI Queloz, grant number 327127). J.d.W. and MIT gratefully acknowledge financial support from the Heising-Simons Foundation, Dr. and Mrs. Colin Masson and Dr. Peter A. Gilman for Artemis, the first telescope of the SPECULOOS network situated in Tenerife, Spain. This work is supported by the Swiss National Science Foundation (PP00P2-163967, PP00P2-190080 and the National Centre for Competence in Research PlanetS). This work has received fund from the European Research Council (ERC)
 under the European Union's Horizon 2020 research and innovation programme (grant agreement n$^\circ$ 803193/BEBOP), from the MERAC foundation, and from the Science and Technology Facilities Council (STFC; grant n$^\circ$ ST/S00193X/1) and from the ERC/UKRI Frontier Research Guarantee programme (EP/Z000327/1/CandY). TRAPPIST is funded by the Belgian Fund for Scientific Research (Fond National de la Recherche Scientifique, FNRS) under the grant PDR T.0120.21, with the participation of the Swiss National Science Fundation (SNF). Data have been collected from the ESO la Silla Observatory in Chile. SPECULOOS and TRAPPIST also acknowledge support from the F.R.S.-FNRS (PDR 35292916 and 40028002). MG and EJ are F.R.S.-FNRS Research Directors. 
A. S postdoctoral fellowship is funded by F.R.S.-FNRS research project ID 40028002 (Detection and Study of Rocky Worlds)

This material is based upon work supported by the National Aeronautics and Space Administration under Agreement No.\ 80NSSC21K0593 for the program ``Alien Earths''.
The results reported herein benefited from collaborations and/or information exchange within NASA’s Nexus for Exoplanet System Science (NExSS) research coordination network sponsored by NASA’s Science Mission Directorate.
This material is based upon work supported by the European Research Council (ERC) Synergy Grant under the European Union’s Horizon 2020 research and innovation program (grant No.\ 101118581---project REVEAL).
Visiting Astronomer at the Infrared Telescope Facility, which is operated by the University of Hawaii under contract 80HQTR24DA010 with the National Aeronautics and Space Administration.
Funding for KB was provided by the European Union (ERC AdG SUBSTELLAR, GA 101054354).
YGMC was supported by UNAM-PAPIIT-IG101224. 
Author F.J.P acknowledges financial support from the Severo Ochoa grant CEX2021-001131-S funded by MCIN/AEI/10.13039/501100011033 and 
Ministerio de Ciencia e Innovación through the project PID2022-137241NB-C43.
This work makes use of observations from the LCOGT network. Part of the LCOGT telescope time was granted by NOIRLab through the Mid-Scale Innovations Program (MSIP). MSIP is funded by NSF.
This research has made use of the Exoplanet Follow-up Observation Program (ExoFOP; DOI: 10.26134/ExoFOP5) website, which is operated by the California Institute of Technology, under contract with the National Aeronautics and Space Administration under the Exoplanet Exploration Program.
Funding for the TESS mission is provided by NASA's Science Mission Directorate. KAC acknowledges support from the TESS mission via subaward s3449 from MIT.
This work is partly supported by JSPS KAKENHI Grant Numbers JP24H00017, JP24K00689, and JSPS Bilateral Program Number JPJSBP120249910.
This paper is based on observations made with the MuSCAT instruments, developed by the Astrobiology Center (ABC) in Japan, the University of Tokyo, and Las Cumbres Observatory (LCOGT). MuSCAT3 was developed with financial support by JSPS KAKENHI (JP18H05439) and JST PRESTO (JPMJPR1775), and is located at the Faulkes Telescope North on Maui, HI (USA), operated by LCOGT. MuSCAT4 was developed with financial support provided by the Heising-Simons Foundation (grant 2022-3611), JST grant number JPMJCR1761, and the ABC in Japan, and is located at the Faulkes Telescope South at Siding Spring Observatory (Australia), operated by LCOGT.
The paper is based on observations made with the Kast spectrograph on the Shane 3m telescope at Lick Observatory. A major
upgrade of the Kast spectrograph was made possible through generous gifts from the Heising-Simons Foundation and William and
Marina Kast. We acknowledge that Lick Observatory sits on the unceded ancestral homelands of the Chochenyo and Tamyen Ohlone
peoples, including the Alson and Socostac tribes, who were the
original inhabitants of the area that includes Mt. Hamilton.
T.D. acknowledges support from the McDonnell Center for the Space Sciences at Washington University in St. Louis.
Some of the observations in this paper made use of the High-Resolution Imaging instrument Zorro and were obtained under Gemini LLP Proposal Number: GN/S-2021A-LP-105. Zorro was funded by the NASA Exoplanet Exploration Program and built at the NASA Ames Research Center by Steve B. Howell, Nic Scott, Elliott P. Horch, and Emmett Quigley. Zorro was mounted on the Gemini South telescope of the international Gemini Observatory, a program of NSF’s OIR Lab, which is managed by the Association of Universities for Research in Astronomy (AURA) under a cooperative agreement with the National Science Foundation. on behalf of the Gemini partnership: the National Science Foundation (United States), National Research Council (Canada), Agencia Nacional de Investigación y Desarrollo (Chile), Ministerio de Ciencia, Tecnología e Innovación (Argentina), Ministério da Ciência, Tecnologia, Inovações e Comunicações (Brazil), and Korea Astronomy and Space Science Institute (Republic of Korea).
B.-O. D. and YGMC acknowledge support from the Swiss National Science Foundation (IZSTZ0$\_$216537)
\section*{Data Availability}
TESS data products are available via the MAST portal at https: \url{//mast.stsci.edu/portal/Mashup/Clients/Mast/Portal.html}
Follow-up observations (photometry, high-resolution imaging data) are available on ExoFOP or on request.



\bibliographystyle{mnras}
\bibliography{6716} 

\vspace{1cm}
\noindent $^{1}$School of Physics \& Astronomy, University of Birmingham, Edgbaston, Birmingham B15 2TT, UK\\
$^{2}$Department of Astrophysics, University of Oxford, Denys Wilkinson Building, Keble Road, Oxford OX1 3RH, UK\\
$^{3}$Magdalen College, University of Oxford, Oxford OX1 4AU, UK\\
$^{\text{4}}$Astrobiology Research Unit, Université de Liège, 19C Allée du 6 Août, 4000 Liège, Belgium\\
$^{\text{5}}$Department of Earth, Atmospheric and Planetary Science, Massachusetts Institute of Technology, 77 Massachusetts Avenue, Cambridge, MA 02139, USA \\
$^{\text{6}}$Kavli Institute for Astrophysics and Space Research, Massachusetts Institute of Technology, Cambridge, MA 02139, USA \\
$^{\text{7}}$Instituto de Astrofísica de Canarias (IAC), Calle Vía Láctea s/n, 38200, La Laguna, Tenerife, Spain \\
$^{\text{8}}$Department of Astronomy \& Astrophysics, UC San Diego, La Jolla, CA 92039, USA \\
$^{\text{9}}$Center for Astrophysics | Harvard \& Smithsonian, 60 Garden Street, Cambridge, MA, 02138, USA \\
$^{\text{10}}$NASA Ames Research Center, Moffett Field, CA 94035, USA \\
$^{\text{11}}$Instituto de Astrofísica de Andalucía (IAA-CSIC), Glorieta de la Astronomía s/n, 18008 Granada, Spain \\
$^{\text{12}}$Department of Physics \& Astronomy, Vanderbilt University, 6301 29 \\
$^{\text{13}}$Department of Physics, Engineering and Astronomy, Stephen F. Austin State University, 1936 North St, Nacogdoches, TX 75962, USA \\
$^{\text{14}}$Universidad Nacional Autónoma de México, Instituto de Astronomía, AP 70-264, Ciudad de México, 04510, México \\
$^{\text{15}}$NASA Exoplanet Science Institute, IPAC, California Institute of Technology, Pasadena, CA 91125 USA \\
$^{\text{16}}$ Department of Physics and McDonnell Center for the Space Sciences, Washington University, St. Louis, MO 63130, USA \\
$^{\text{17}}$Center for Space and Habitability, University of Bern,
Gesellschaftsstrasse 6, 3012, Bern, Switzerland \\
$^{\text{18}}$Flatiron Institute Center for Computational Astrophysics, New York, NY 10010 \\
$^{\text{19}}$Komaba Institute for Science, The University of Tokyo, 3-8-1 Komaba, Meguro, Tokyo 153-8902, Japan \\
$^{\text{20}}$European Space Agency (ESA), European Space Research and Technology Centre (ESTEC), Keplerlaan 1, 2201 AZ Noordwijk, The Netherlands \\
$^{\text{21}}$Space sciences, Technologies \& Astrophysics Research (STAR) Institute, University of Liège, Liège, Belgium. \\
$^{\text{22}}$Institute for Particle Physics and Astrophysics, ETH Zürich, Wolfgang-Pauli-Strasse 2, 8093 Zürich, Switzerland \\
$^{\text{23}}$Department of Physics and Astronomy, The University of North Carolina at Chapel Hill, Chapel Hill, NC 27599-3255, USA \\
$^{\text{24}}$Cavendish Laboratory, JJ Thomson Avenue, Cambridge CB3 0HE, UK \\
$^{\text{25}}$Astrobiology Center, 2-21-1 Osawa, Mitaka, Tokyo 181-8588, Japan \\

\appendix

\section{Archival Images}

\begin{figure*} 
  \centering
  \subfloat[Archival images for TOI-6716.]{\includegraphics[width=0.9\linewidth]{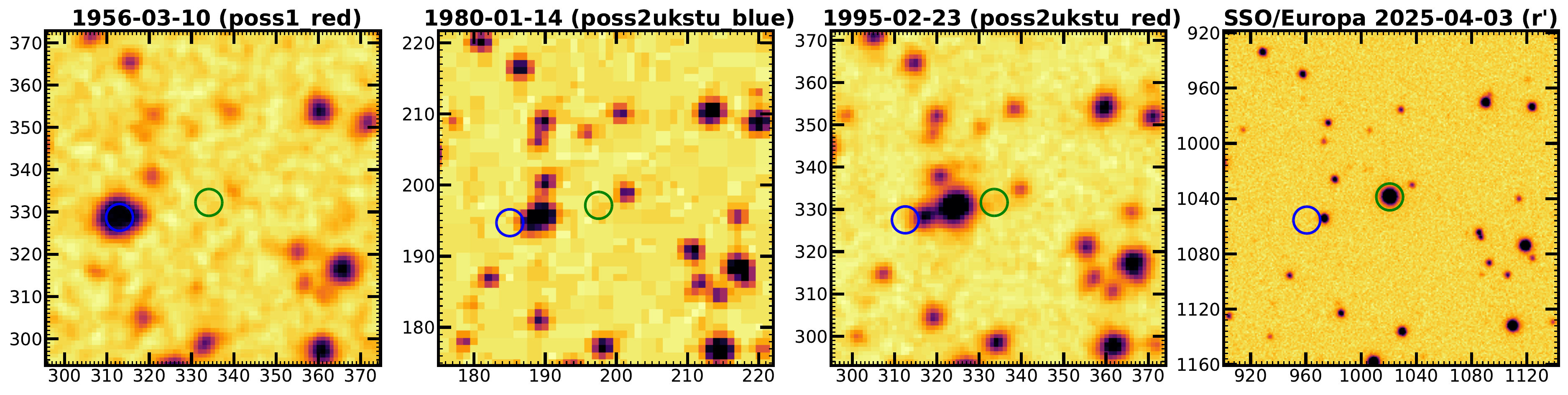}}\label{fig:archival_6716}\quad 
  \subfloat[Archival images for TOI-7384.]{\includegraphics[width=0.9\linewidth]{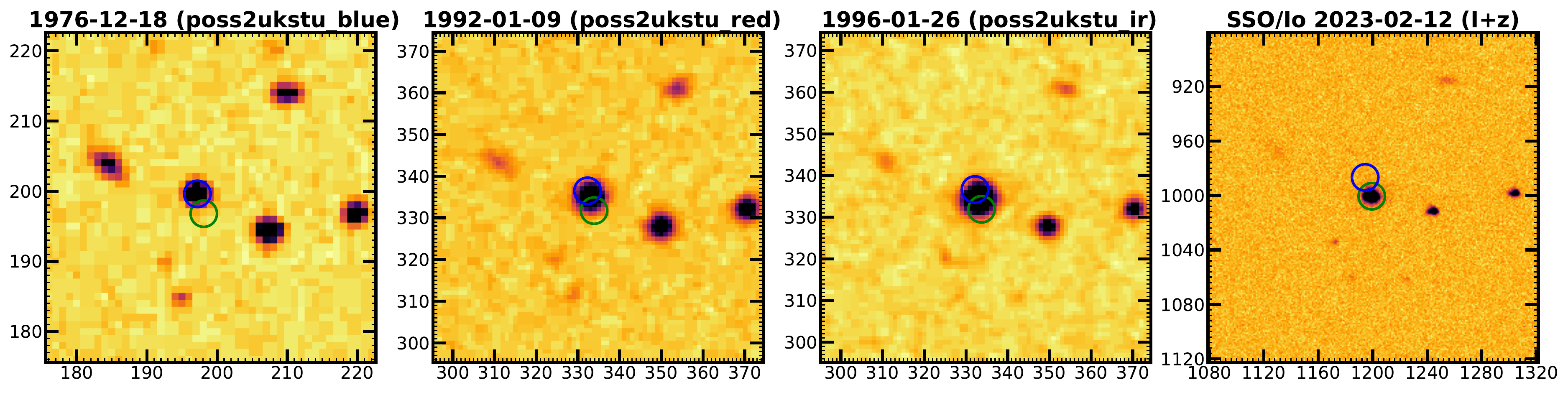}} \label{fig:archival_7384}
  \caption{Archival images for TOI-6716 \textit{(top)} and TOI-7384 (\textit{bottom)}. Blue circles show the position of the star in the earliest image, and green circles show the position of the star in the most recent images. TOI-6716 has shifted by 22.2" from 1956 to 2025, and TOI-7384 by 5.2" from 1976 to 2023. These images show that there is no background star blending with either of our target stars, and thus we conclude the transit event could not have have happened on a blended star.}
  \label{fig:archival}
\end{figure*}

\section{High-resolution images - Zorro}

\begin{figure*} 
  \centering
  \subfloat[TOI-6716.]{\includegraphics[width=\columnwidth]{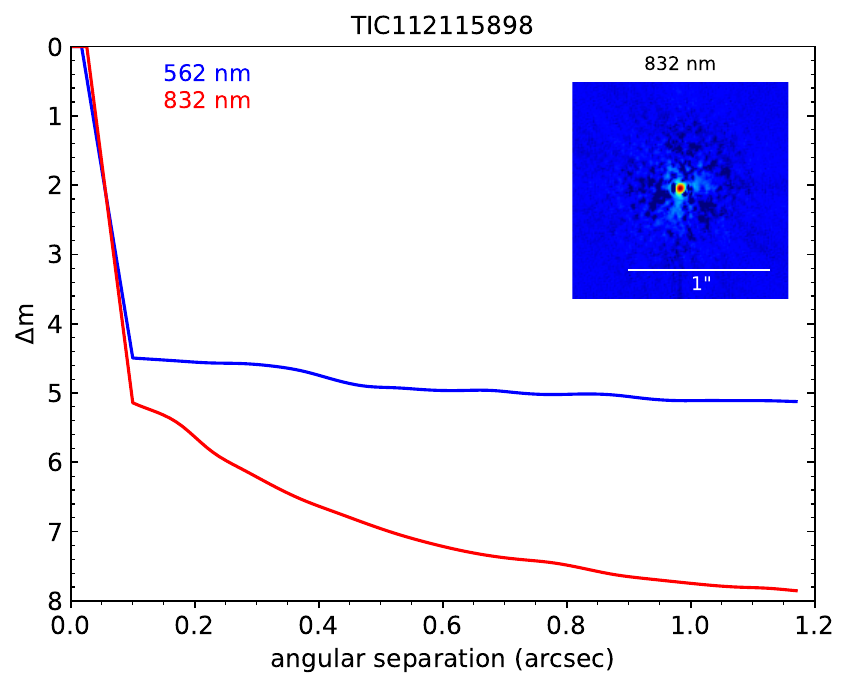}}\quad 
  \subfloat[TOI-7384.]{\includegraphics[width=\columnwidth]{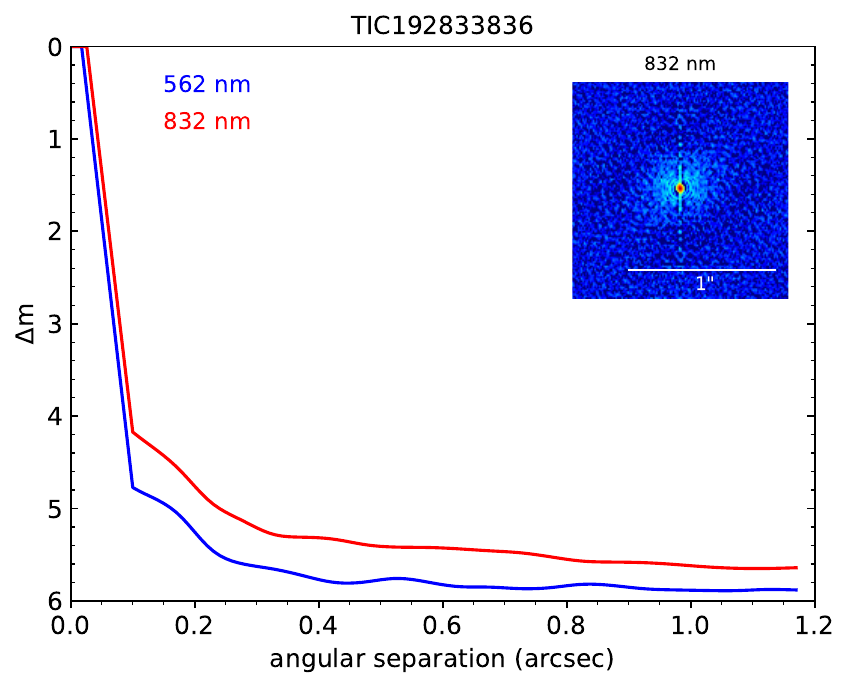}} 
  \caption{The figure shows $5\sigma$ magnitude contrast curves in both filters as a function of the angular separation out to 1.2 arcsec. The inset shows the reconstructed 832 nm image of TOI-6716 \textit{(left)} and TOI-7384 \textit{(right)} with a 1 arcsec scale bar. Both stars were found to have no close companions from the diffraction limit (0.02”) out to 1.2 arcsec to within the magnitude contrast levels achieved.}
  \label{fig:zorro}
\end{figure*} 

\section{High-resolution images - SOAR}

\begin{figure}
    \centering
    \includegraphics[width=\columnwidth]{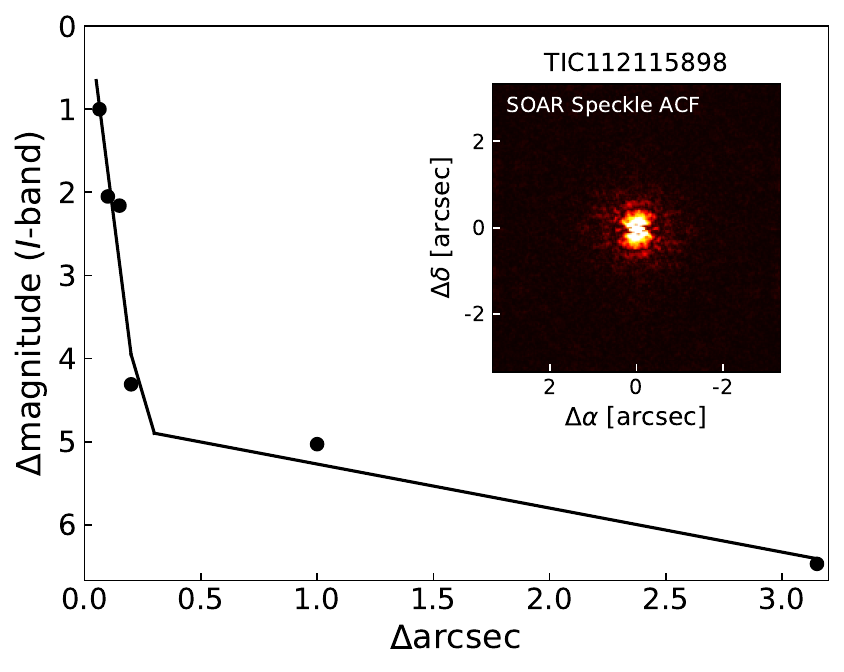}
    \caption{The 5~$\sigma$ detection sensitivity to nearby companions for the SOAR speckle observation of TOI-6716 as a function of separation from the target star and magnitude difference with respect to the target. Inset is the speckle auto-correlation function from the observation.}
    \label{fig:SOAR_6716}
\end{figure}

\section{Priors for Global Photometric Analysis}

\begin{table} 
\centering
\caption{Priors used in achromatic eccentric fit for TOI-6716\,b and TOI-7384\,b}
\begin{tabular}{@{}ccccc@{}} 
\midrule \midrule
Parameter & \multicolumn{2}{c}{{Fit Parameterization}} \\ 
\hline 
 & \textit{6716\,b} & \textit{7384\,b} \\
\hline 
$R_b / R_\star$ & $\mathcal{U}(0.005, 0.05)$ & $\mathcal{U}(0.05, 0.15)$  \\
$(R_\star + R_b) / a_b$ & $\mathcal{U}(0.02, 0.06)$ & $\mathcal{U}(0.03, 0.05)$  \\
$\cos{i_b}$ &  $\mathcal{U}(0.0, 0.05)$ & $\mathcal{U}(0.0, 0.1)$\\
$T_{0;b}$ ($\mathrm{BJD}$) &  $\mathcal{U}(2458495.1, 2458495.21)$ & $\mathcal{U}(2459683.5, 2459683.6)$\\
$P_b$ ($\mathrm{d}$) & $\mathcal{U}(4.71, 4.72)$ & $\mathcal{U}(6.23, 6.24)$\\
$\sqrt{e_b} \cos{\omega_b}$ & $\mathcal{U}(-1, 1)$ & $\mathcal{U}(-1, 1)$ \\
$\sqrt{e_b} \sin{\omega_b}$ & $\mathcal{U}(-1, 1)$ & $\mathcal{U}(-1, 1)$ \\
$q_{1; \mathrm{TESS}}$ & $\mathcal{N}_{0,1}(0.296, 0.05)$ & $\mathcal{N}_{0,1}(0.285, 0.005)$ \\
$q_{2; \mathrm{TESS}}$ & $\mathcal{N}_{0,1}(0.295, 0.05)$ & $\mathcal{N}_{0,1}(0.304, 0.005)$\\
$q_{1; \mathrm{{i'}}}$ &  $\mathcal{N}_{0,1}(0.400, 0.05)$ & $\mathcal{N}_{0,1}(0.343, 0.005)$ \\
$q_{2; \mathrm{{i'}}}$ &  $\mathcal{N}_{0,1}(0.288, 0.05)$ & $\mathcal{N}_{0,1}(0.309, 0.005)$ \\
$q_{1; \mathrm{{g'}}}$ & $\mathcal{N}_{0,1}(0.763, 0.05)$ & $\mathcal{N}_{0,1}(0.688, 0.005)$ \\
$q_{2; \mathrm{{g'}}}$ & $\mathcal{N}_{0,1}(0.333, 0.05)$ & $\mathcal{N}_{0,1}(0.359, 0.005)$ \\
$q_{1; \mathrm{{r'}}}$ & $\mathcal{N}_{0,1}(0.667, 0.05)$ & $\mathcal{N}_{0,1}(0.597, 0.005)$ \\
$q_{2; \mathrm{{r'}}}$ & $\mathcal{N}_{0,1}(0.361, 0.05)$ & $\mathcal{N}_{0,1}(0.377, 0.005)$ \\
$q_{1; \mathrm{{z_s}}}$ & $\mathcal{N}_{0,1}(0.294, 0.05)$ & - \\
$q_{2; \mathrm{{z_s}}}$ & $\mathcal{N}_{0,1}(0.262, 0.05)$ & - \\
$q_{1; \mathrm{{I+z'}}}$ & - & $\mathcal{N}_{0,1}(0.266, 0.005)$ \\ 
$q_{2; \mathrm{{I+z'}}}$ & - & $\mathcal{N}_{0,1}(0.288, 0.005)$ \\ 
$q_{1; \mathrm{{z'}}}$ & - & $\mathcal{N}_{0,1}(0.128, 0.005)$ \\ 
$q_{2; \mathrm{{z'}}}$ & - & $\mathcal{N}_{0,1}(0.287, 0.005)$ \\ 
$q_{1; \mathrm{spirit}}$ & - & $\mathcal{N}_{0,1}(0.133, 0.005)$ \\ 
$q_{2; \mathrm{spirit}}$ & - & $\mathcal{N}_{0,1}(0.288, 0.005)$ \\
$\ln{\sigma_\mathrm{all}}$ & $\mathcal{U}(-30, 1)$ & $\mathcal{U}(-15, 0)$  \\
\end{tabular} \label{tab:priors}
\end{table}


\bsp	
\label{lastpage}
\end{document}